\documentclass[12pt,preprint]{aastex}

\shorttitle{Outflows in $\rho$ Ophiuchi}
\shortauthors{Zhang \& Wang}

\begin{document}
\title{Outflows in $\rho$ Ophiuchi as Seen with the Spitzer Infrared Array Camera}
\author{Miaomiao Zhang and Hongchi Wang}
\affil{Purple Mountain Observatory, Chinese Academy of Sciences, Nanjing 210008, PR China}
\email{miaomiao@pmo.ac.cn}

\begin{abstract}
Using the IRAC images from the {\it Spitzer c2d} program, we have made a survey of mid-infrared outflows in the $\rho$ Ophiuchi molecular cloud. Extended objects that have prominent emission in IRAC channel 2 (4.5 \micron) compared to IRAC channel 1 (3.6 \micron) and stand out as green objects in the three-color images (3.6 \micron \ in blue, 4.5 \micron \ in green, 8.0 \micron \ in red) are identified as mid-infrared outflows. As a result, we detected 13 new outflows in the $\rho$ Ophiuchi molecular cloud that have not been previously observed in optical or near-infrared. In addition, at the positions of previously observed HH objects or near-infrared emission, we detected 31 mid-infrared outflows, among which seven correspond to previously observed HH objects and 30 to near-infrared emission. Most of the mid-infrared outflows detected in the $\rho$ Ophiuchi cloud are concentrated in the L1688 dense core region. In combination with the survey results for Young Stellar Objects (YSOs) and millimeter and sub-millimeter sources, the distribution of mid-infrared outflows in the $\rho$ Ophiuchi molecular complex hints a propagation of star formation in the cloud in the direction from the northwest to the southeast as suggested by previous studies of the region.
\end{abstract}


\section{Introduction}
Mass outflows play an essential role in the process of star formation and have been found to be ubiquitous in various stages of star formation \citep{arc07,bal07}. The specific angular momentum of a star-forming molecular core is about four orders of magnitude times higher than that of a typical T Tauri star \citep{bod95}. One way to transfer the excessive angular momentum from the star-forming cores so that the circumstellar material can be accreted onto the central stars is through mass outflows \citep{sha07}. Mass outflows from YSOs have been observed at different wavelengths. In the visual, high velocity jets and HH objects with a typical velocity of 100 - 300 km s$^{-1}$ trace material ejected by the star or shocked surrounding medium \citep{rei01}. Near-infrared molecular hydrogen emission features, which usually have velocities of several tens km s$^{-1}$, trace H$_2$ gas in the jets or in the entrained surrounding medium \citep{eis00}. In the millimeter wavelength, high velocity CO gas with velocity in the range from a few to ten km s$^{-1}$ have been observed \citep{wu04}. The high velocity CO gas probes the swept or entrained medium. The different velocity ranges and excitation conditions for HH objects, near-infrared molecular hydrogen emission, and CO outflows indicate the transfer of energy and momentum from the high velocity material to the surrounding medium.

Shocked gas in outflows has abundant atomic and molecular hydrogen line emission in the mid-infrared. Hydrodynamic simulations of outflows from YSOs by \citet{sr05} have shown that outflows have strong H$_2$ line emission in all the IRAC four bands. In fact, mid-infrared observations with IRAC aboard of {\it Spitzer} have discovered a lot of new outflows \citep{mer08} and plenty new features of previously known outflows \citep{nor04,smith06,tei08}, showing that {\it Spitzer} IRAC imaging is a powerful tool in the survey of outflows, particular for those that are deeply embedded in clouds due to the much less extinction in the mid-infrared compared to the optical and near-infrared wavelengths. The {\it Spitzer} legacy program {\it c2d} performed a complete imaging of five nearby star-forming regions with the IRAC and MIPS instruments \citep{eva03,eva09}, which provides a valuable data set to search for mid-infrared outflows in these regions. In this paper we present the results of a mid-infrared outflow survey of the $\rho$ Ophiuchi molecular cloud based on the {\it c2d} archive images.

The $\rho$ Ophiuchi molecular cloud is one of the closest star-forming regions at a distance of about 130 pc \citep{wil08}. It is intermediate in star formation activity compared to the isolated star formation in the Taurus cloud and the rich clusters in the Orion cloud. The large scale structure of the $\rho$ Ophiuchi cloud has been revealed by extensive molecular $^{13}$CO line mapping \citep{lor89a,lor89b}. Millimeter and submillimeter surveys of the central part of the cloud have revealed the clumpy structure of the cloud and have found about 55 starless cores with masses in the range of 0.02 to 6.3 M$_\sun$ \citep{motte98,johnstone00,you06}. The cloud has been surveyed for young stars at wavelengths from X-ray \citep{montmerle83,imanishi01,ozawa05}, visual \citep{wilking87,wilking05}, near-infrared \citep{gre92,allen02}, mid- and far-infrared \citep{you86,wilking01,pad08}, to millimeter and submillimeter \citep{sta06,andrews07}. Based on these surveys, \citet{wil08} compiled a list of 316 young stars in L1688, the main cloud of the $\rho$ Ophiuchi complex. About 46 HH objects \citep{wilking97,wu02,phe04}, 119 H$_2$ emission features \citep{gro01,gom03,kha04}, 16 high velocity CO outflows \citep{bontemps96,bussmann07} have been observed in the $\rho$ Ophiuchi complex. For a summary of the $\rho$ Ophiuchi complex we refer to the recent review by \citet{wil08}

\section{The Data}
\subsection{The Spitzer Data}

\begin{figure}
\epsscale{1.0}
\plotone{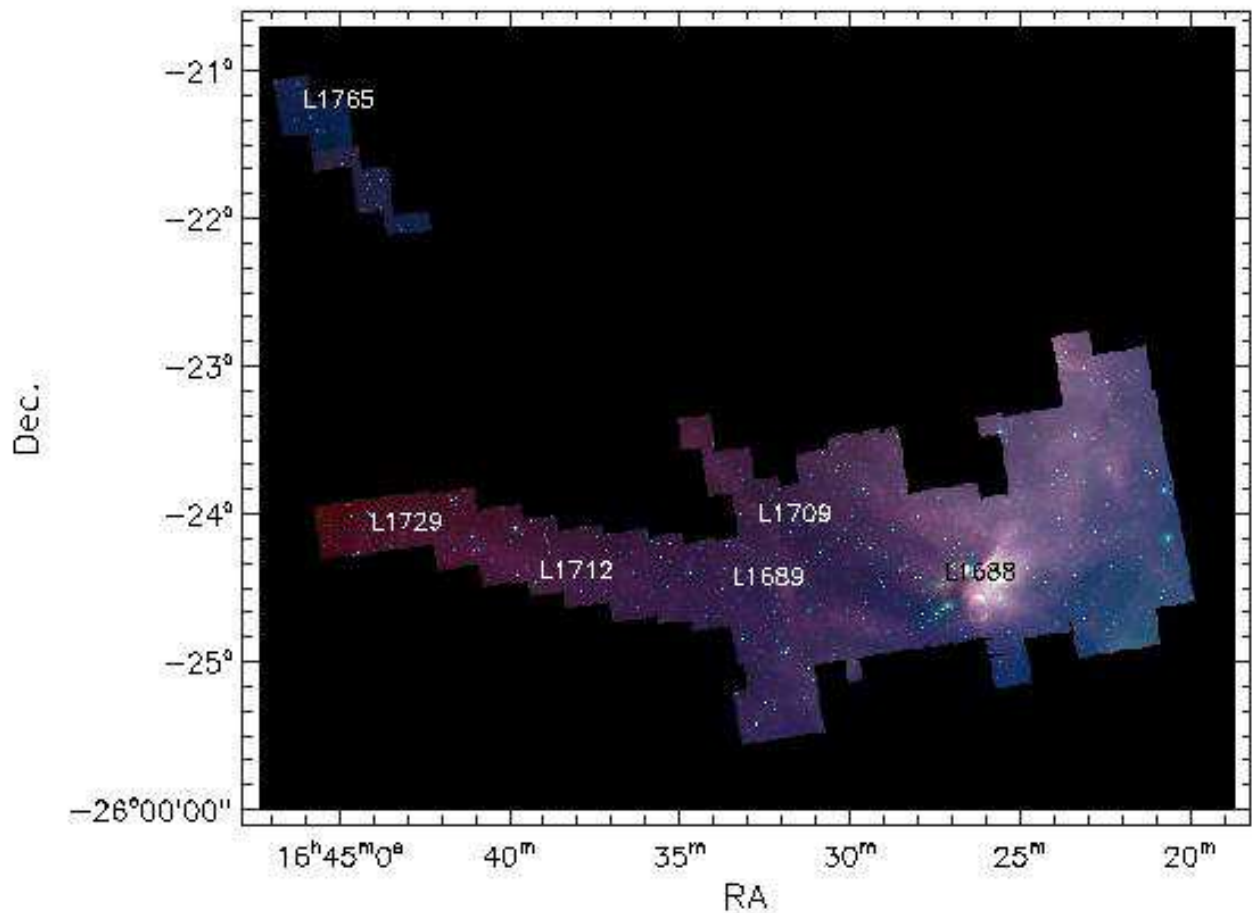}
\caption{Three-color image of the $\rho$ Ophiuchi molecular clouds with IRAC band 1 (3.6\micron, blue), IRAC band 2 (4.5\micron, green) and IRAC band 4 (8.0\micron, red). We marked the main Lynds cloud in the region \citep{lyn62}.\label{fig1}}
\end{figure}

The IRAC observations of $\rho$ Ophiuchi were conducted on 2004 March 28-30 and 2005 September 15 and 18. These observations are part of the Spitzer legacy program {\it c2d} \citep{eva03} and the program ID is 177. Fig.~\ref{fig1} shows the three-color image of the area mapped with IRAC which is about 8.0 deg$^2$. This area is included in the region mapped with MIPS (see Fig.1 of \citealt{pad08}). Each region in Fig.~\ref{fig1} was individually mapped in two epochs with an exposure time of 12 seconds (24 seconds in total) and one map was made in the high dynamic range (HDR) mode, which added a short exposure time of 0.6 seconds \citep{eva03}. Six off-cloud fields were observed in order to sample the background source counts.

We used the final mosaic images and source catalogs of the fourth delivery of IRAC and MIPS data. The data were processed with {\it Spitzer} Science Center pipeline S13 and calibrated by {\it c2d} team (details can be found in the Delivery Document\footnote{http://data.spitzer.caltech.edu/popular/c2d/20071101\_enhanced\_v1/Documents/\\c2d\_del\_document.pdf}).
In the final mosaic images bright source artifacts have been removed. However, some instrumental effects still survive. For example, optical ghosts produced by internal reflections within the filters (labelled with ``gh" in the subsequent figures) and stray lights produced by illuminating sources off the edges of the detector arrays (labelled in the subsequent figures with ``Stray light") remain in the final mosaic images (see the IRAC data handbook\footnote{http://ssc.spitzer.caltech.edu/irac/dh/iracdatahandbook3.0.pdf} for details).

A catalog of young stellar object candidates in the $\rho$ Ophiuchi, YSOc, is also delivered by the {\it c2d} team. \citet{eva09} improved over this catalog. They removed some suspect sources, added known sources and data obtained at other wavelengths, and calculated additional quantities such as bolometric temperature and bolometric luminosity. Their list makes a relatively pure YSO candidate sample which includes 292 YSOs. We used their list of YSOs in this paper. Two schemes for YSO classification were used in \citet{eva09}. One is the spectral index that was obtained using a least-squares fit to all photometry between 2 \micron \ and 24 \micron, and the criteria for classification are from \citet{gre94}. The other scheme used the calculation of bolometric temperature \citep{chen95,dunham08}. Using the calculated spectral indices or bolometric temperatures, 292 YSOs in the cloud can be classified into 4 classes: 27 in Class I or Class 0, 44 in the flat-spectrum Class, 179 in Class II, and 42 in Class III. Note especially that there are
three Class 0 sources in Ophiuchus, IRAS 16293-2422, MMS 126, and VLA 1623-243, which were distinguished from Class I sources on the basis of their bolometric temperatures.
\subsection{The FCRAO Data}
To analyze the distribution of detected outflows in the $\rho$ Ophiuchi molecular complex, we used the $^{13}$CO 1-0 (110.201GHz) emission data from the FCRAO 14 m telescope at New Salem, MA, U.S.A. The data were obtained during 2002-2005 with the On-the-Fly (OTF) mapping method \citep{rid06} and are part of the {\it COMPLETE}\footnote{http://cfa-www.harvard.edu/COMPLETE} survey of star-forming regions. These $^{13}$CO observations cover 10.0 deg$^2$ of the $\rho$ Ophiuchi molecular complex and the FCRAO Half Power Beam Width (HPBW) measures 44\arcsec. The $^{13}$CO integrated intensity maps that have been re-gridded to 23\arcsec were used. In our $^{13}$CO emission contour maps, the units of integrated intensity is antenna temperature (T$_{A}^*$) in Kelvin.

\subsection{Identification of IRAC Mid-infrared Outflows}
IRAC channels cover a relatively broad wavelength range from 3.6 \micron \ to 8.0 \micron. In this wavelength range there are many H$_2$ lines and atomic and ionic fine-structure lines. \citet{ros00} have shown near- and mid-infrared spectra of the brightest H$_2$ emission peak of the Orion OMC-1 outflow which are taken with ISO-SWS. The spectra revealed a wealth of molecular lines of H$_2$, CO, and H$_2$O, and a number of H I recombination lines, atomic and ionic fine-structure lines. The IRAC 4.5 \micron \ band contains strong H$_2$ lines ($v = 0-0$, S(9, 10, 11)) and CO bandheads ($v = 1-0$) \citep{rea06}. Like the near-infrared H$_{2}$ emission at 2.12 \micron, the emission of these lines can be excited by the interaction of protostellar outflows with the ambient medium. \citet{sr05} presented synthetic {\it Spitzer} IRAC band maps from hydrodynamic simulations of processing  protostellar jets and they predicted that the integrated H$_{2}$ line luminosity is greatest in IRAC band 2 (4.5\micron), about an order of magnitude higher than those in other IRAC bands (see Table 3 in \citealt{sr05}). Observations by {\it Spitzer} IRAC have in general confirmed this prediction \citep{nor04,jor06}. According to \citet{nor04}, the remaining emission in the {\it Spitzer} IRAC band 2 image after the subtraction of the emission in band 1 is mainly from shocked H$_{2}$ and CO gas. Using the three-color images constructed from IRAC bands 1, 2, and 4., \citet{cyg08} compiled a catalog of outflow candidates with the {\it Spitzer} Galactic Legacy Infrared Mid-Plane Survey Extraordinaire (GLIMPSE) data. They identified the Extended Green Objects (EGOs) in the three-color images as outflow candidates. In this paper, we use the difference image constructed from IRAC channel 2 (4.5 \micron) minus channel 1 (3.6 \micron) to search for mid-infrared outflows. We use the three-color (3.6 \micron \ in blue, 4.5 \micron \ in green, 8.0 \micron \ in red) images to confirm that the identified outflows are EGOs.

Images covering the common sky area of the $\rho$ Ophiuchi cloud in the four IRAC bands were trimmed from the downloaded {\it c2d} archive mosaic images. The trimmed images were then rotated so that the image north direction is top and the east is to the left. For the subtraction of emission in band 1 from that in band 2, we selected 80 bright stars in the images that are roughly evenly distributed over the sky and are not saturated in either band 1 or band 2. The average flux ratio in band 1 to band 2 of these 80 stars is used to adjust the flux level of stars in band 1 to that in band 2. The resultant difference images were visually inspected to search for mid-infrared outflows. We selected extended sources that have significant excess at 4.5\micron \ as mid-infrared outflows. Toward the core of L1688, several surveys for HH objects and H$_{2}$ emission have been carried out \citep{wu02,gom03,kha04} and 46 HH objects, including components of HH objects, and 119 H$_{2}$ emission features have been reported. We examined whether there are mid-infrared counterparts of these HH objects and H$_{2}$ emission features at their corresponding positions in the IRAC images.

\section{Results}
We discovered 13 IRAC mid-infrared outflows. Their coordinates and figure reference are list in Table~\ref{tab1}. The likely exciting sources for these outflows and their Lada classes are given in Table~\ref{tab2}. Note that IRAS 16293-2422 is classified as a Class 0 source based on its bolometric temperature \citep{eva09}. We also identified 31 mid-infrared counterparts of previously known HH objects and H$_{2}$ 2.12 \micron \ emission features. The coordinates of these mid-infrared counterparts of the previously known outflows, together with the corresponding HH objects and H$_2$ 2.12 \micron \ emission features, are presented in Table~\ref{tab3}. In this paper we designate all the detected IRAC mid-infrared outflows as EGOs.

The EGOs are numbered by order in right ascension. If the overall morphology of a group of knots suggests physical association among them, the knots are considered as parts of one EGO and the individual knots are distinguished with additional letters to the EGO numbers. On the other hand, if the knots in a region are relatively separated from each other and the relationship among them is not clear, each knot is treated as an individual EGO. Apparent sub-structures in a knot are labelled with additional numerals, such as EGO 21b1.

\begin{figure}
\epsscale{1.0}
\plotone{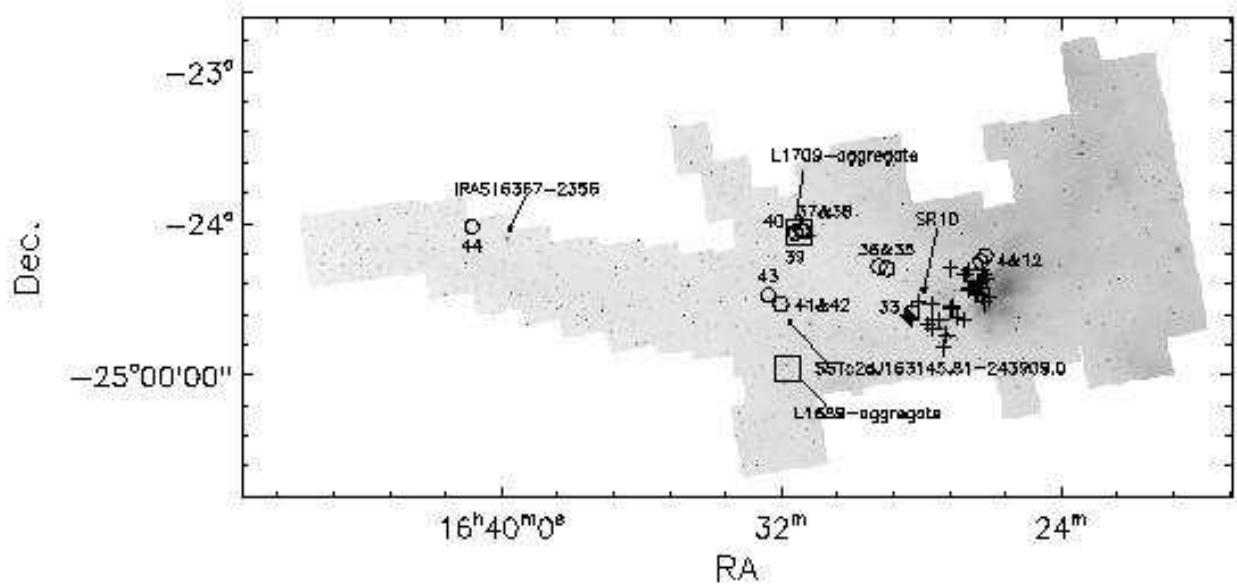}
\caption{The portion of the $\rho$ Ophiuchi cloud which includes all the IRAC mid-infrared outflows. The grey scale image is the IRAC channel 2 (4.5\micron) image. The newly detected outflows are marked with circles and the counterparts of previously known HH objects and H$_{2}$ flows are labelled with pluses. The open squares represent the L1689 and L1709 aggregates \citep{pad08}. We also labelled three YSOs with filled squares for the convenience of subsequent discussion (see Section 3.1). SR sources are from \citet{str49}.\label{fig2}}
\end{figure}

\begin{figure}
\plotone{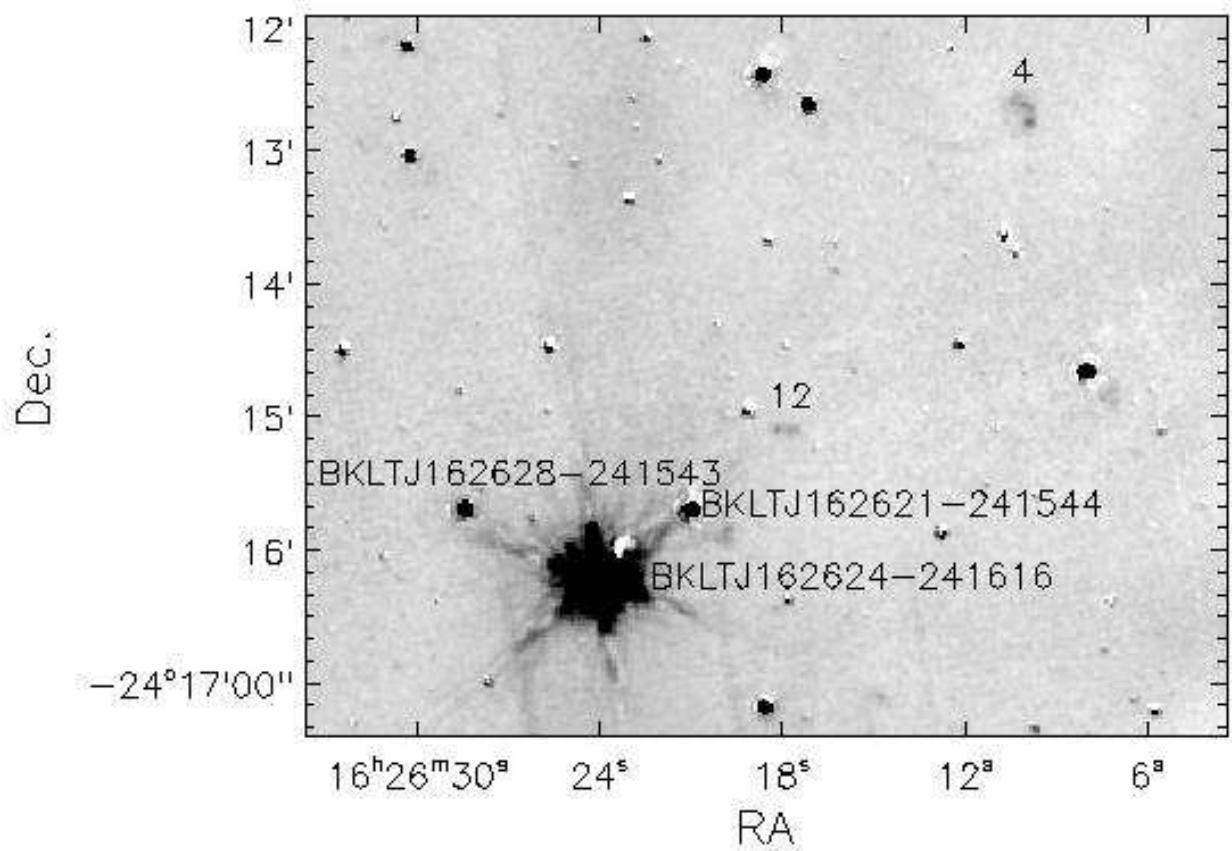}
\caption{The region of EGO 04 and EGO 12. The image is the difference image of IRAC channel 2 (4.5 \micron) minus channel 1 (3.6 \micron). The BKLT sources are from \citet{bar97}. \label{fig3}}
\end{figure}

\begin{figure}
\plotone{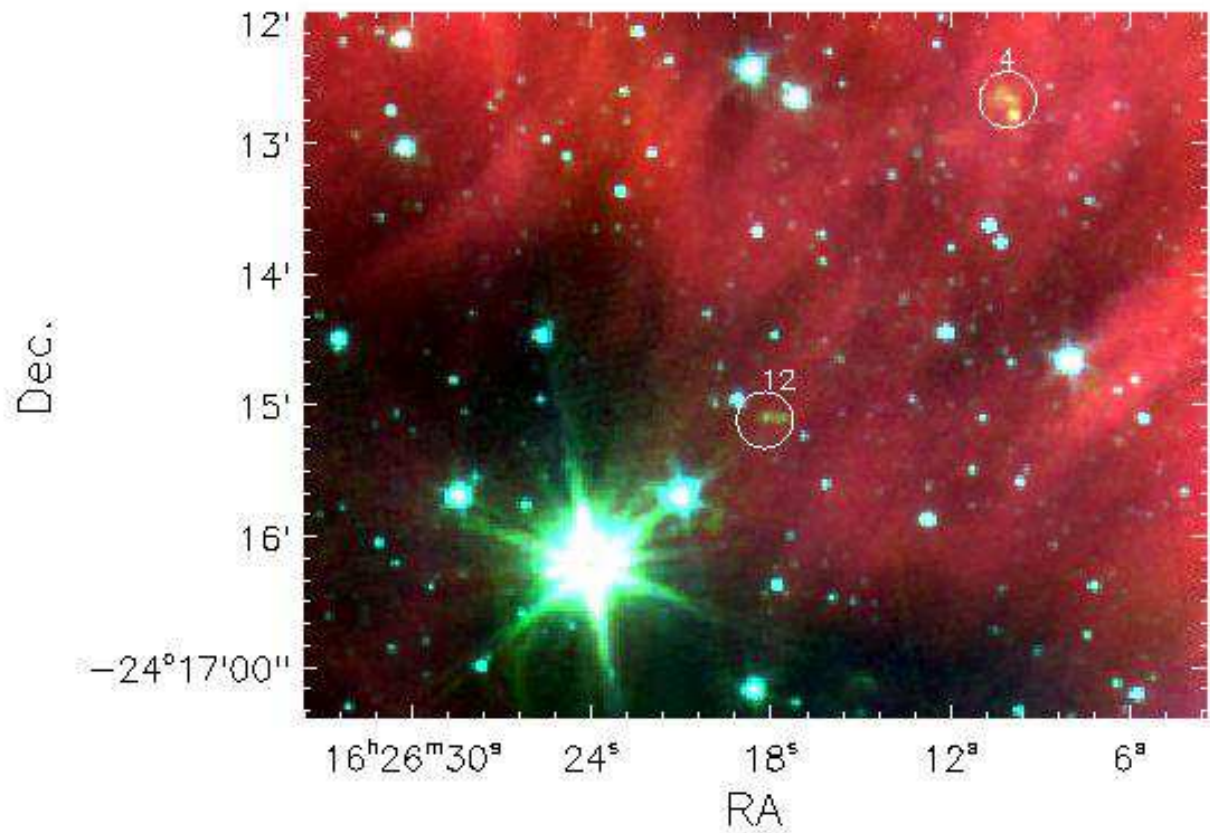}
\caption{Three-color image of the EGO 04 and EGO 12 region. The image uses IRAC channel 1 (3.6 \micron, blue), channel 2 (4.5 \micron, green) and channel 4 (8.0 \micron, red). The EGOs identified in the region are marked.\label{fig4}}
\end{figure}

\begin{figure}
\plotone{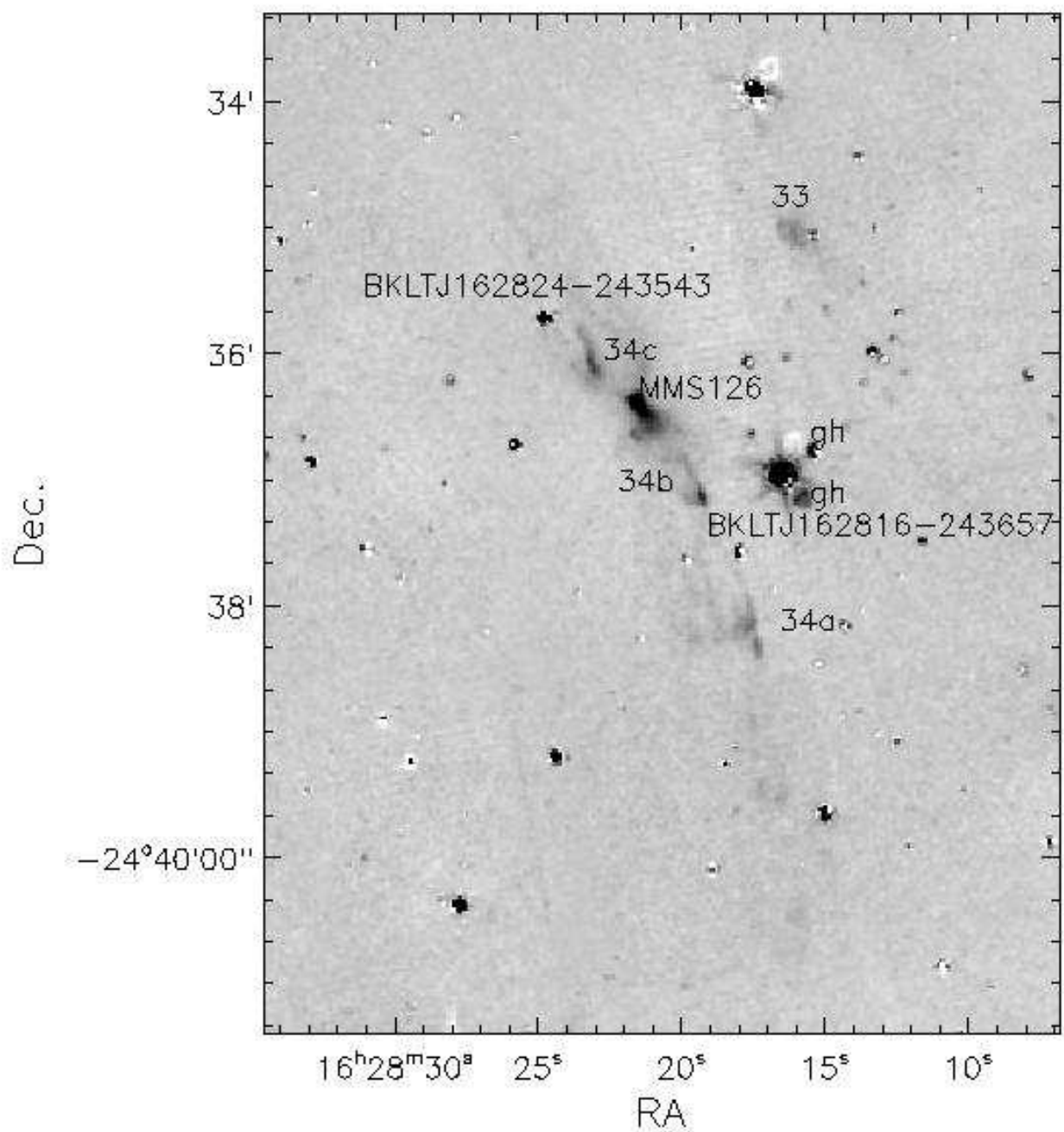}
\caption{The difference image of the EGO 33 and EGO 34 region. Filter ghosts are marked with ``gh". MMS126 (Class 0, \citealt{sta06}) and BKLT sources \citep{bar97} are marked. \label{fig5}}
\end{figure}

\begin{figure}
\plotone{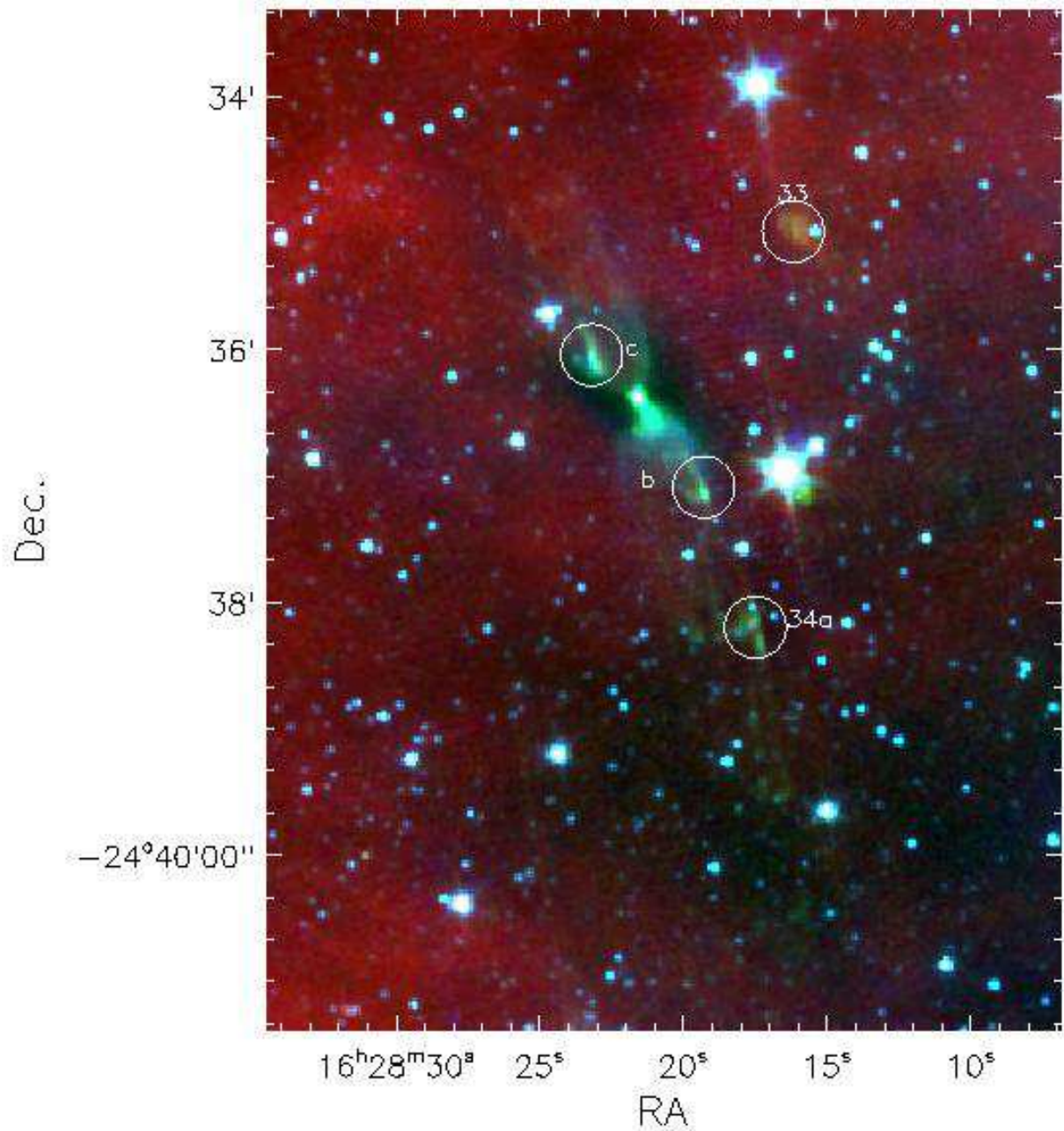}
\caption{Three-color image of the EGO 33 and EGO 34 region. Others are the same as in Fig.~\ref{fig4}.\label{fig6}}
\end{figure}

\begin{figure}
\epsscale{1.0}
\plotone{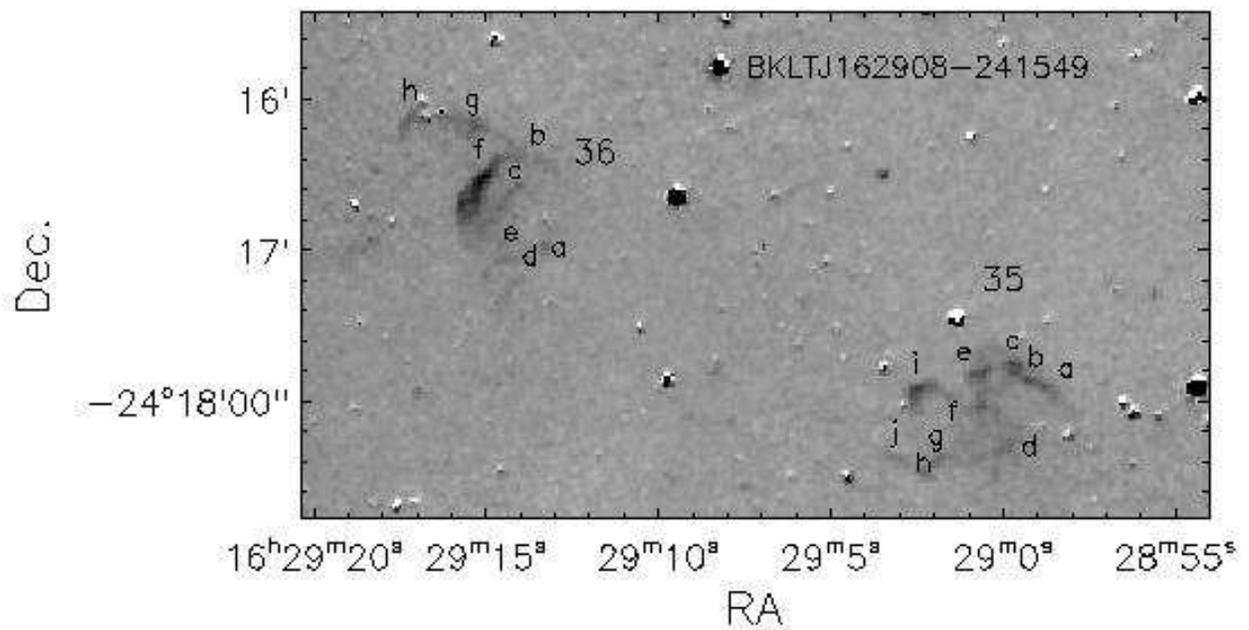}
\caption{Region of EGOs 35 and 36. Others are the same as in Fig.~\ref{fig3}\label{fig7}}
\end{figure}

\begin{figure}
\epsscale{1.0}
\plotone{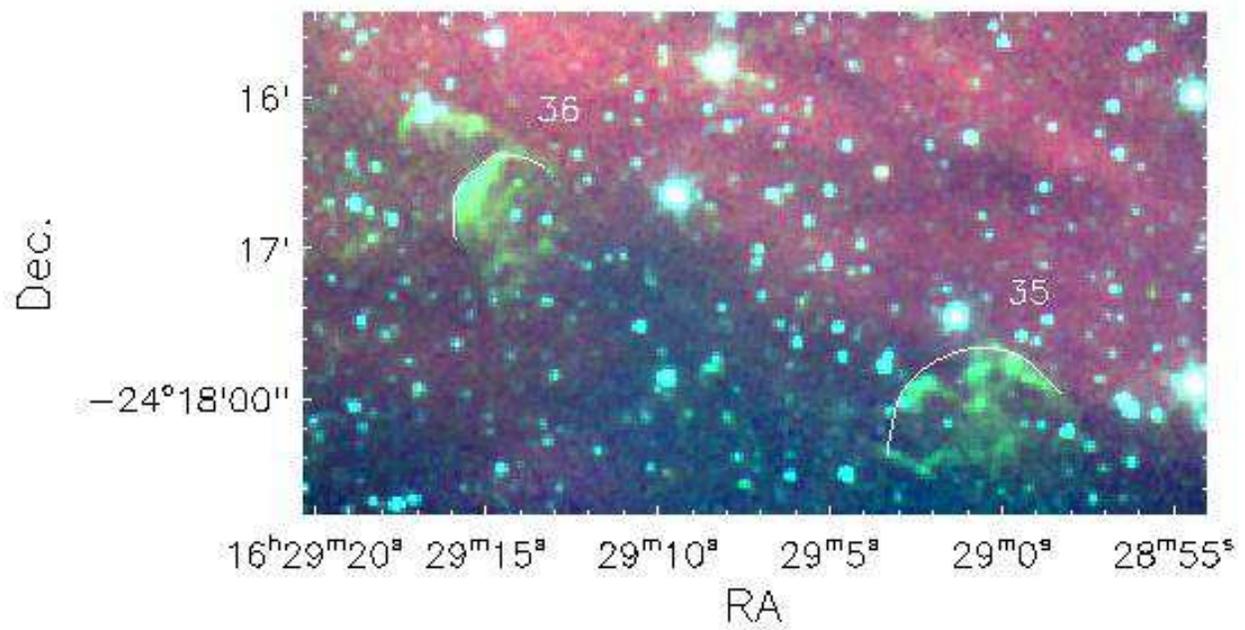}
\caption{Three-color image of the region of EGOs 35 and 36. Others are the same as in Fig.~\ref{fig4}. Note that the bow shock morphology of EGOs 35 and 36 is outlined with the white lines.\label{fig8}}
\end{figure}

\begin{figure}
\epsscale{1.0}
\plotone{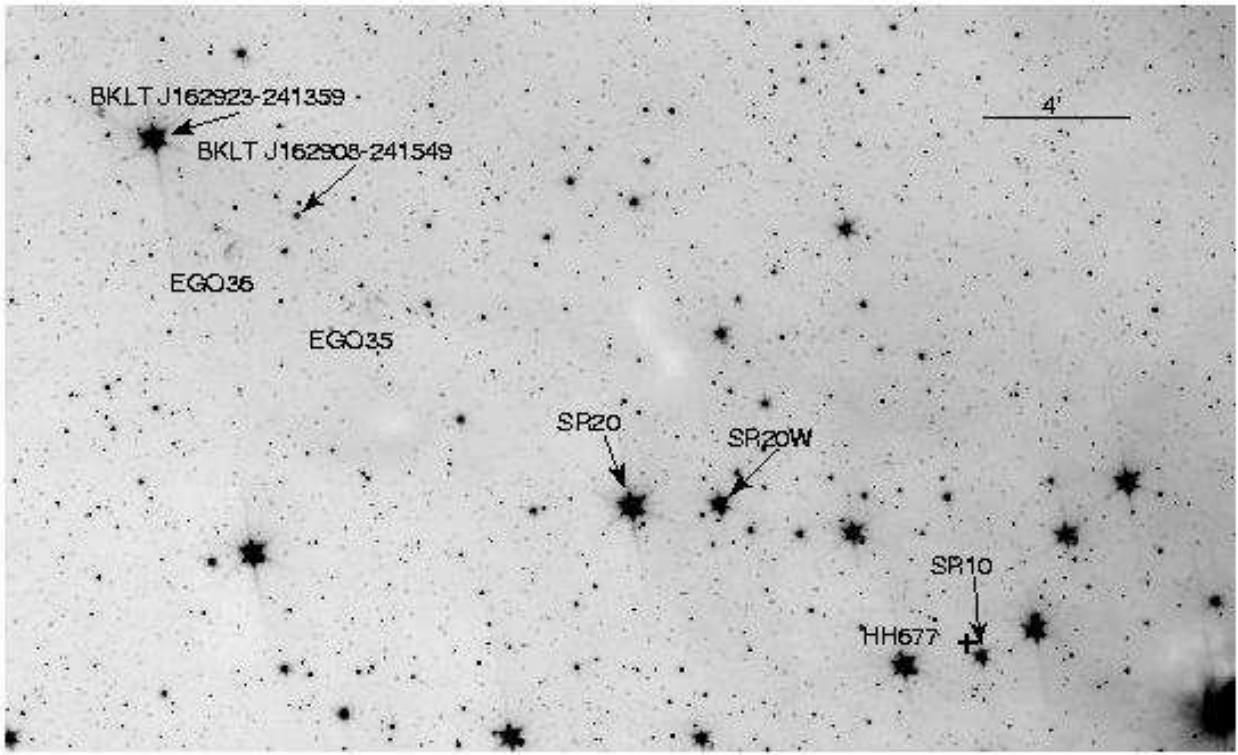}
\caption{A large view of the EGOs 35 and 36 region. YSOs in this region are marked.\label{fig9}}
\end{figure}

\begin{figure}
\epsscale{1.0}
\plotone{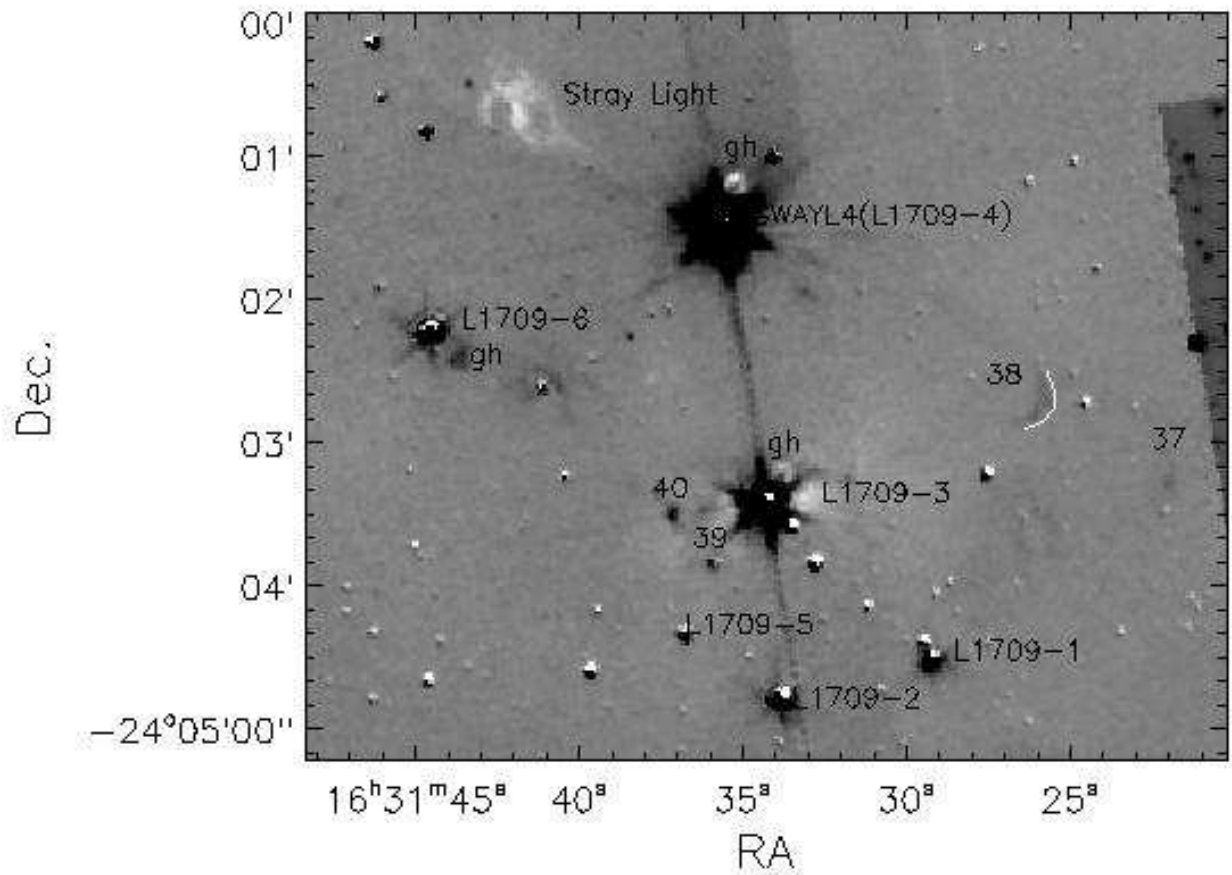}
\caption{The region of EGOs 37-40.  The members of the L1709 YSO aggregate, L1709 1-6 \citep{pad08}, are labelled. Filter ghosts and stray lights are marked. The bow shock morphology of EGO 38 is outlined with a white line. Others are the same as in Fig.~\ref{fig3}\label{fig10}}
\end{figure}

\begin{figure}
\epsscale{1.0}
\plotone{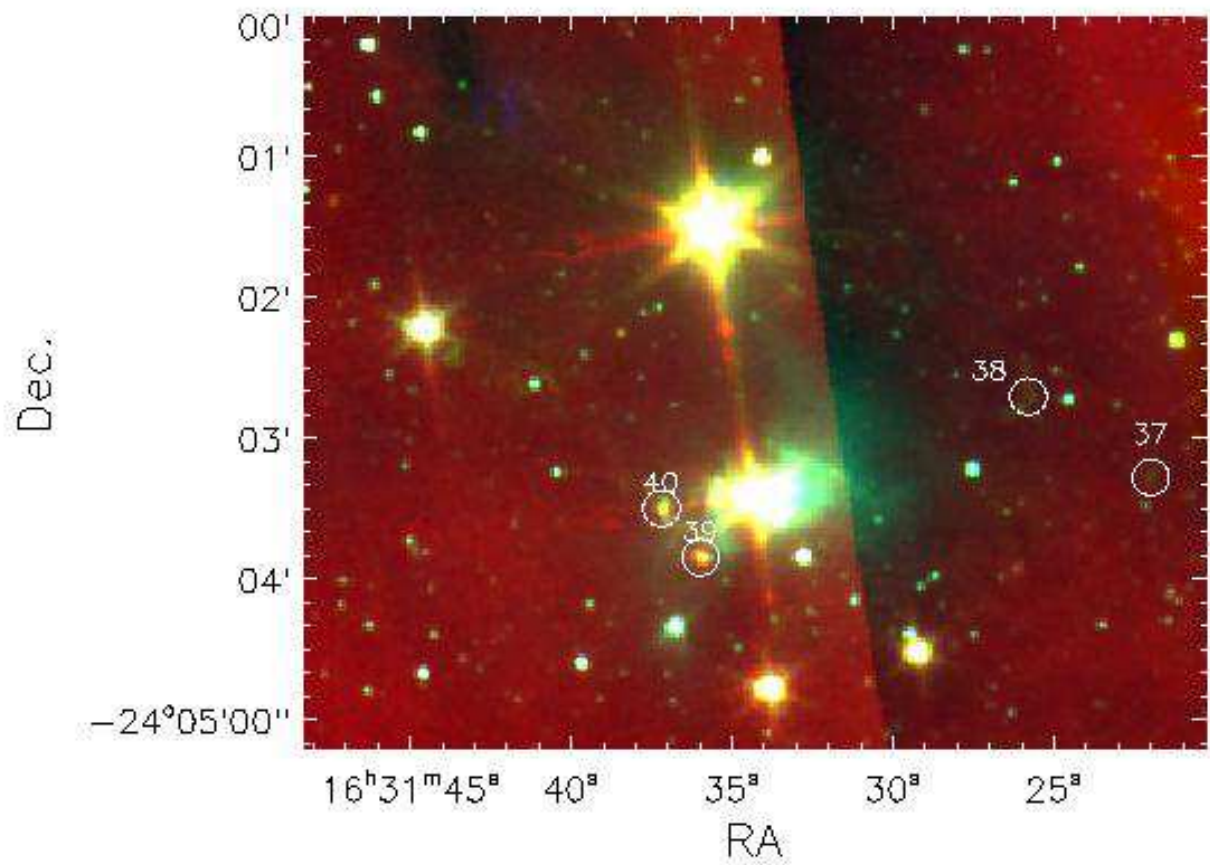}
\caption{Three-color image of the region of EGOs 37-40. Others are the same as in Fig.~\ref{fig4}.\label{fig11}}
\end{figure}

\begin{figure}
\epsscale{1.0}
\plotone{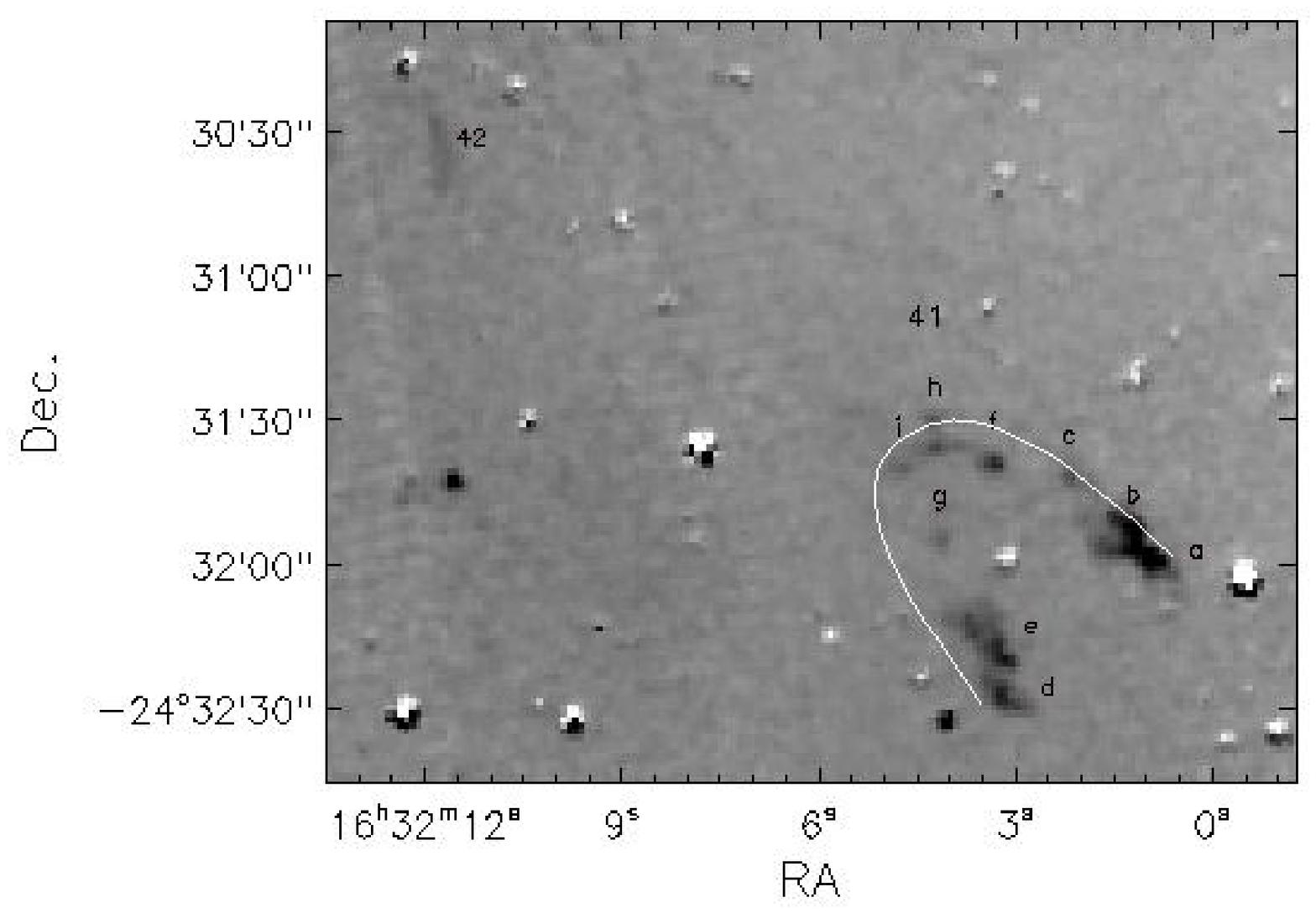}
\caption{The region of EGOs 41 and 42. The bow shock morphology of EGO 41 is outlined with a white line. Others are the same as in Fig.~\ref{fig3}. \label{fig12}}
\end{figure}

The locations of all the mid-infrared outflows identified in the $\rho$ Ophiuchi cloud are shown in Fig.~\ref{fig2}. The grey scale image in Fig.~\ref{fig2} is the IRAC channel 2 (4.5 \micron) image. The outflows newly discovered in this work are marked with circles and the counterparts of known flows are labelled with pluses. Two YSO aggregates in the region, L1689- and L1709-aggregate \citep{pad08}, are marked with open squares. The identified mid-infrared outflows are located mainly in three regions, i.e., the L1688 core, the IRAS 16293-2422 region, and the L1709 YSO aggregate region. The images of the newly discovered IRAC mid-infrared outflows are shown in Figs.~\ref{fig3}-\ref{fig17}, and the images of mid-infrared counterparts of HH objects and H$_2$ emission features are presented in Figs.~\ref{fig18}-\ref{fig40}.

\subsection{New IRAC Mid-infrared Outflows}
Thirteen new mid-infrared outflows have been discovered in our survey. For each mid-infrared outflow, we present the channel 2 (4.5 \micron) minus 1 (3.6 \micron) image and the three-color image.

Figures 3-4 show the region of objects EGO 04 and EGO 12. The distance between the two objects is about 3\arcmin. Each object consists of two knots. EGO 04 and EGO 12 are only detected in IRAC channels 1 (3.6 \micron) and 2 (4.5 \micron). It is brighter in channel 2 (4.5 \micron), therefore, appears as an EGO in the three-color image (Fig.~\ref{fig4}). There are three YSOs to the southeast of EGO 04 and EGO 12, among which we suggest BKLT J162624-241616 (Class II in \citealt{eva09}) as the exciting source of the EGO 04 and EGO 12 outflow, as this source is also detected in the MIPS 24 \micron \ image. The distance from BKLT J162624-241616 to EGO 04 is 4\farcm7 and the distance to EGO 12 is 1\farcm8.

Outflow EGO 33 (Figs. 5 and 6) is a diffuse nebula and is visible in IRAC channels 2 (4.5 \micron), 3 (5.8 \micron), and 4 (8.0 \micron). Mid-infrared counterparts of the known near-infrared outflow in this region, three H$_{2}$ emission features in \citet{kha04}, are also detected in the IRAC images. We designated these mid-infrared counterparts as EGO 34a-c (see Table~\ref{tab3}). EGO 33 appears green and yellow in the three-color image (Fig.~\ref{fig6}), which indicates that EGO 33 may have some emission from dust or PAH. There are two YSOs in this region, MMS126 (Class 0 in \citealt{sta06}) and BKLT J162816-243657 (Class II in \citealt{eva09}). BKLT J162816-243657 is associated with 1.3 mm and mid-infrared emission \citep{and94,gre94}. These YSOs are visible in all IRAC four channels and in the MIPS 24 \micron \ image. We suggest BKLT J162816-243657 as the driving source of EGO 33 on the basis that MMS126 has close association with the known outflow detected in near-infrared and in CO (3-2) emission \citep{sta06}. EGO 33 is unlikely to be associated with MMS 126 because EGO 33 is not coincident with the axis of the known CO outflow. The distance from BKLT J162816-243657 to EGO 33 is about 1\farcm9.

Both EGO 35 and EGO 36 are complexes of emission filaments ( Figs. 7 and 8). Their components are list in Table~\ref{tab1}. EGOs 35 and 36 are visible in IRAC channels 1-3 and are brightest in channel 2 (4.5 \micron). The nearby YSOs are shown in Fig.~\ref{fig9}. BKLT J162908-241549 (Class II in \citealt{eva09}) is the nearest YSO to EGOs 35 and 36. However, the driving source of EGOs 35 and 36 should be located in the direction of the line connecting EGOs 35 and 36 if there is some association between EGOs 35 and 36. Other YSOs in this region are located to the northeast or southwest of EGOs 35 and 36, all roughly aligned with EGOs 35 and 36. The morphology of EGO 35 and EGO 36 is bow-shock-like, with wings extending to the southwest, which suggests that the outflow source should be located to the southwest. HH 677 is about 30\arcsec \ to the northeast of SR 10 and it may be driven by this source \citep{gom03}. We can see that EGOs 35 and 36 and HH 677 are roughly located on a line. Therefore, it is possible that they are physically associated.

EGOs 37-40 are located in the L1709 YSO aggregate \citep{pad08} (Figs.~10-11). EGO 37 is a faint knot and EGO 38 is a faint bow structure and they may have some connection. They are visible in IRAC channels 1 and 2. GWAYL 4, a low-luminosity class I source in \citet{gre94} and classified as a flat spectrum source in \citet{eva09}, is located in the northeastern prolongation of the EGOs 37-38 outflow. The distance from GWAYL 4 to EGO 37 and EGO 38 is about 3\farcm6 and 2\farcm6, respectively. On the basis of the alignment we suggest GWAYL 4 as the exciting source of EGOs 37 and 38. EGOs 39 and 40 are located to the southeast and east of YSO L1709-3, a flat-spectrum source in \citet{eva09}. EGO 39 and EGO 40 are visible in all IRAC four channels and EGO 39 is also visible in the MIPS 24 \micron \ image. The L1709 YSO aggregate is a small cluster of six YSOs and it is difficult at present to identify which YSOs in this cluster as the exciting sources of EGO 39 and EGO 40.

EGO 41 consists of a series of knots, EGO 41a-i, and EGO 42 is a relatively isolated patch (Fig.~\ref{fig12}). Knots EGO 41a-i constitute a bowlike structure facing to northeast. The bowlike structure is visible in all IRAC four channels while EGO 42 is only visible in IRAC channel 2. The symmetry axis of the bowlike structure suggests that the driving source for EGO 41 is located in the NE-SW direction. SSTc2d J163145.81-243909.0, a Class II source \citep{eva09}, is located about 8 \arcmin \ to the southwest of EGO 41 (see Fig.~\ref{fig2}). This source is a possible driving source of EGO 41.

\begin{figure}
\epsscale{1.0}
\plotone{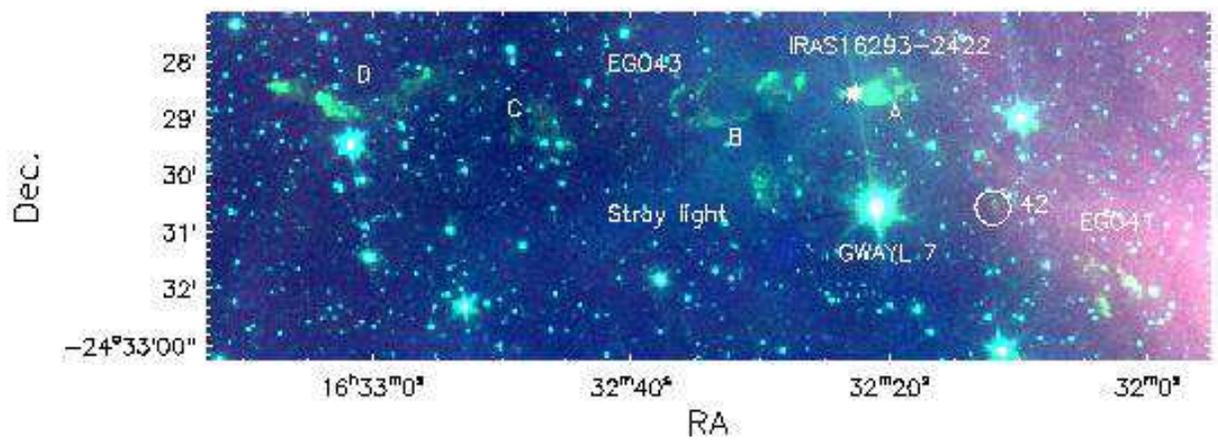}
\caption{Three-color image of the IRAS 16293-2422 region. EGOs in the region, 41-42 and 43A-D, are marked. The GWAYL source is from \citet{gre94}. Others are the same as in Fig.~\ref{fig4}.\label{fig13}}
\end{figure}

\begin{figure}
\epsscale{1.0}
\plotone{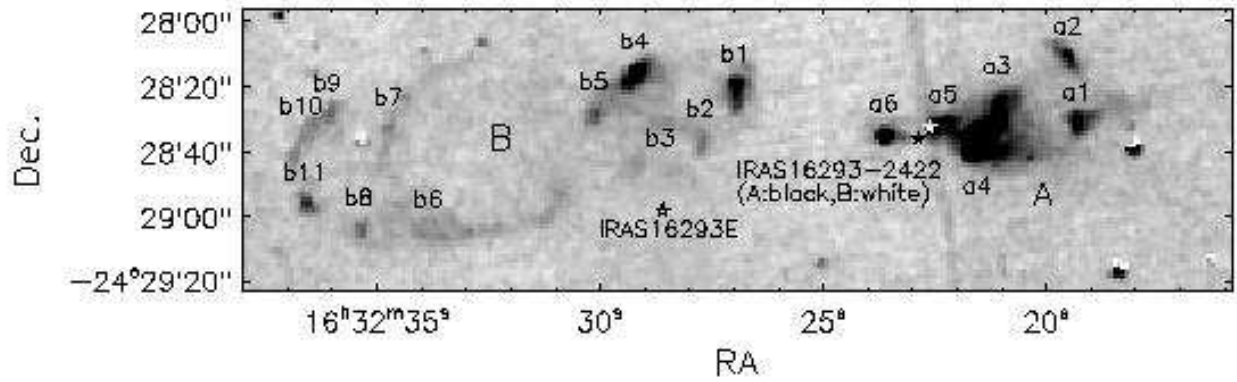}
\caption{Image of the inner portion of the EGO 43 outflow. The locations of IRAS 16293-2422 A, B and IRAS 16293E are marked with pentagrams. Others are the same as in Fig.~\ref{fig3}.\label{fig14}}
\end{figure}

\begin{figure}
\epsscale{1.0}
\plotone{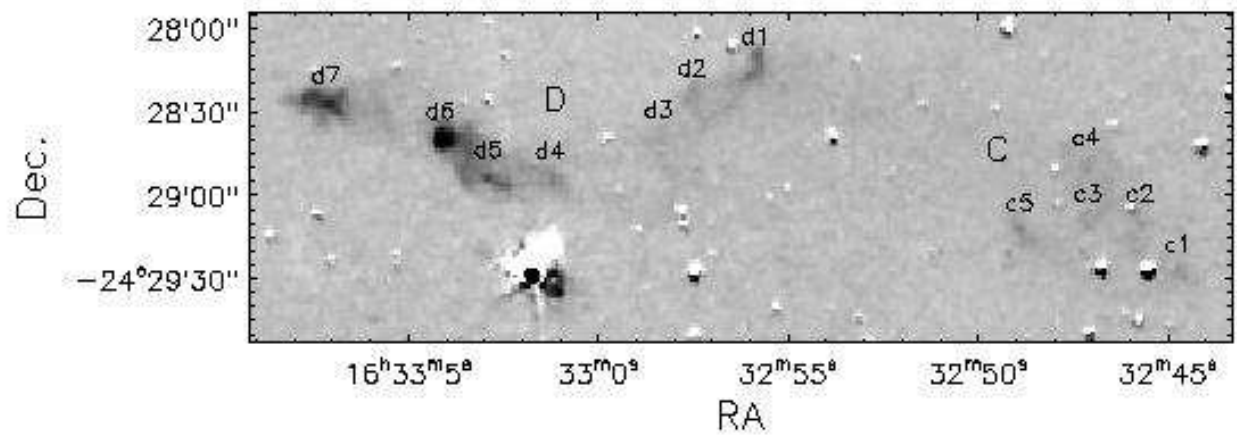}
\caption{Image of the outer portion of the EGO 43 outflow. Others are the same as in Fig.~\ref{fig3}.\label{fig15}}
\end{figure}

\begin{figure}
\epsscale{1.0}
\plotone{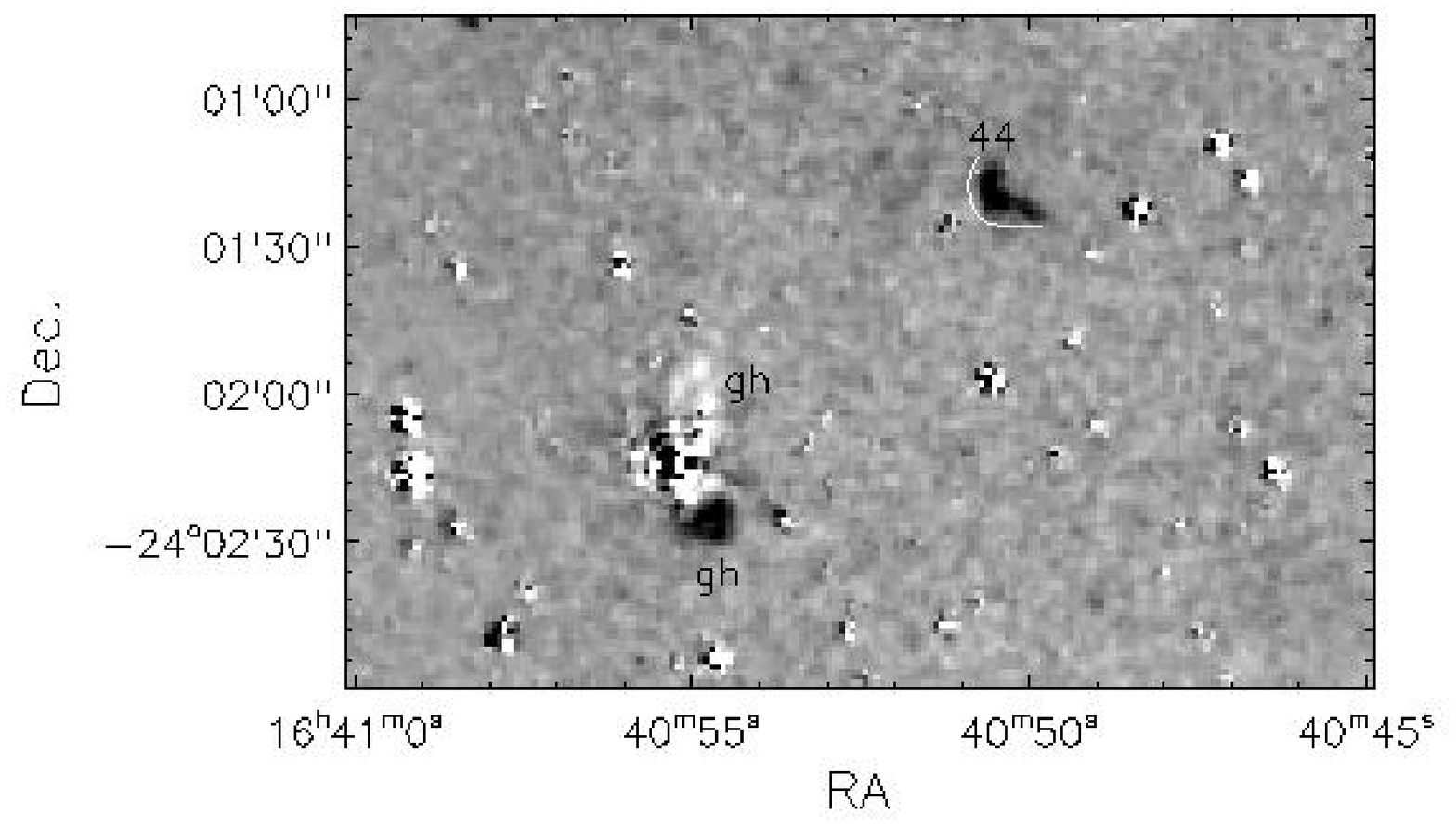}
\caption{The region of EGO 44. The bow shock morphology of EGO 44 is outlined with a white line. Others are the same as in Fig.~\ref{fig3}. \label{fig16}}
\end{figure}

\begin{figure}
\epsscale{1.0}
\plotone{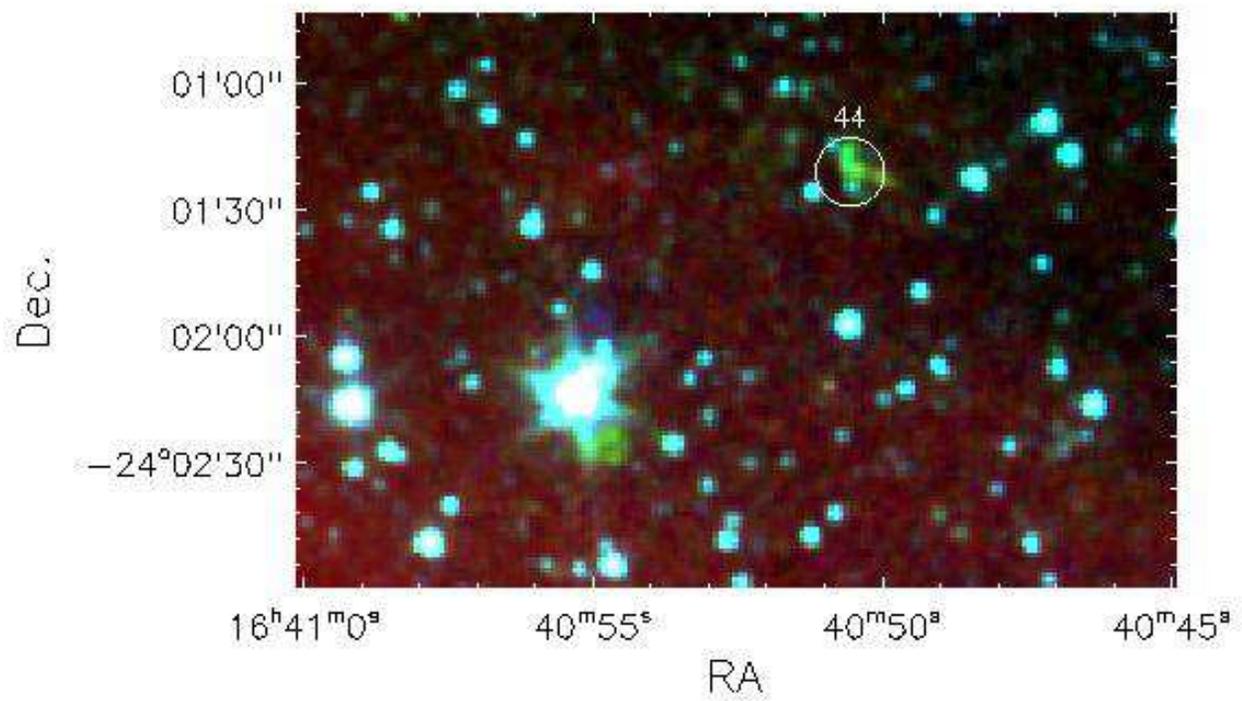}
\caption{Three-color image of the EGO 44 region. Others are the same as in Fig.~\ref{fig4}. \label{fig17}}
\end{figure}

\begin{figure}
\epsscale{1.0}
\plotone{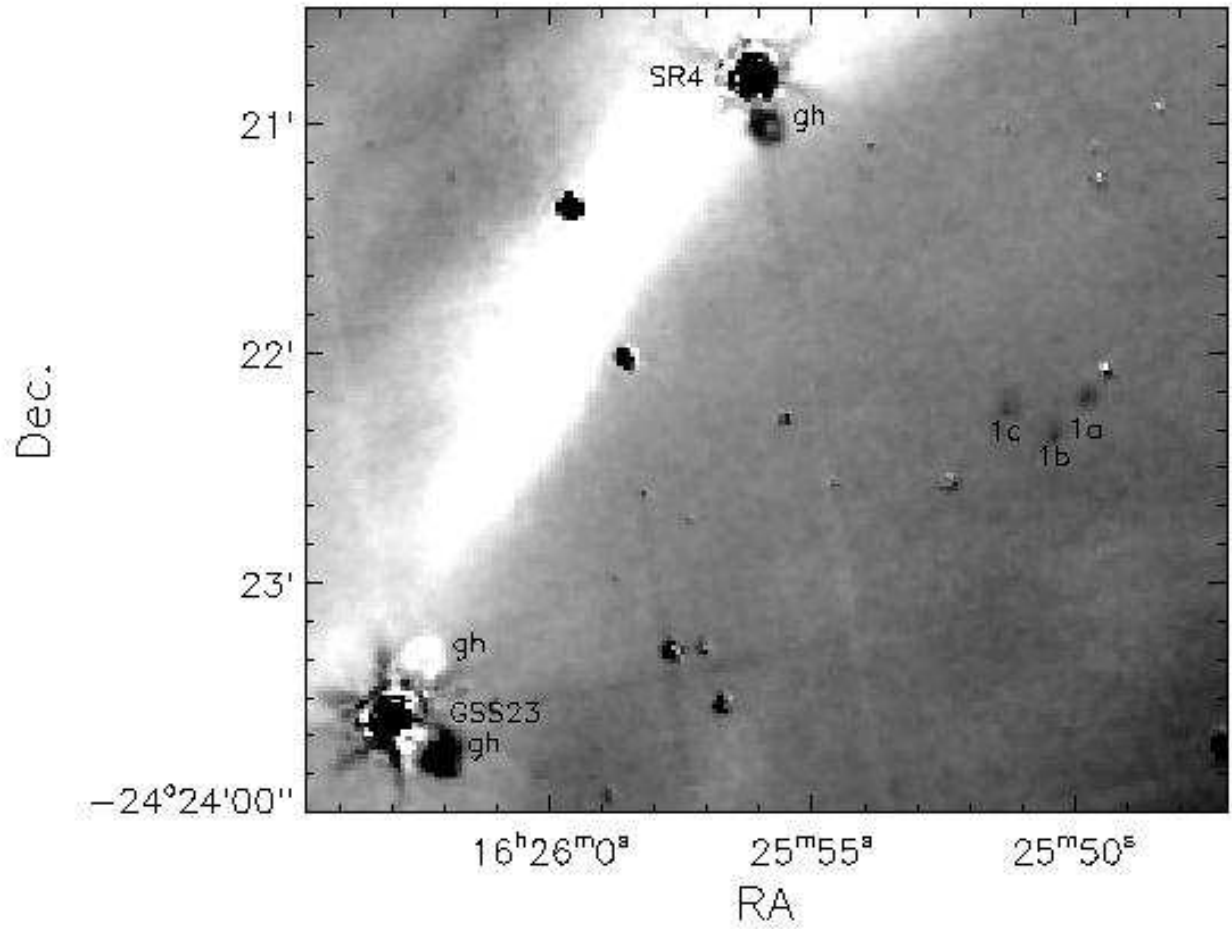}
\caption{The region of EGO 01, a counterpart of H$_{2}$ emission \citep{luc08}. The GSS source is from \citet{gra73} and the SR source is from \citet{str49}. Others are the same as in Fig.~\ref{fig3}. \label{fig18}}
\end{figure}

\begin{figure}
\epsscale{1.0}
\plotone{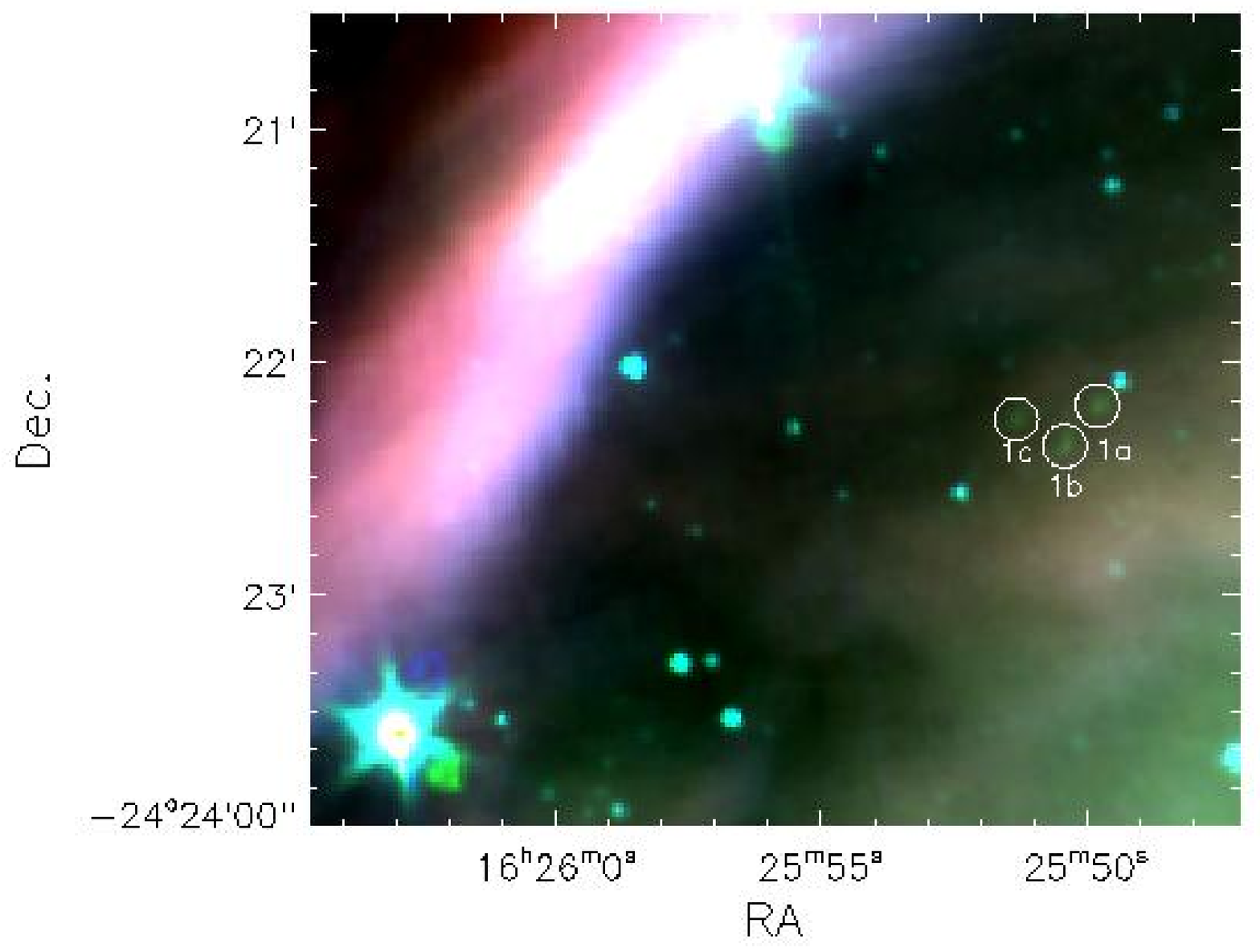}
\caption{The same as Fig.~\ref{fig4}, but for the EGO 01 region.\label{fig19}}
\end{figure}

\begin{figure}
\epsscale{1.0}
\plotone{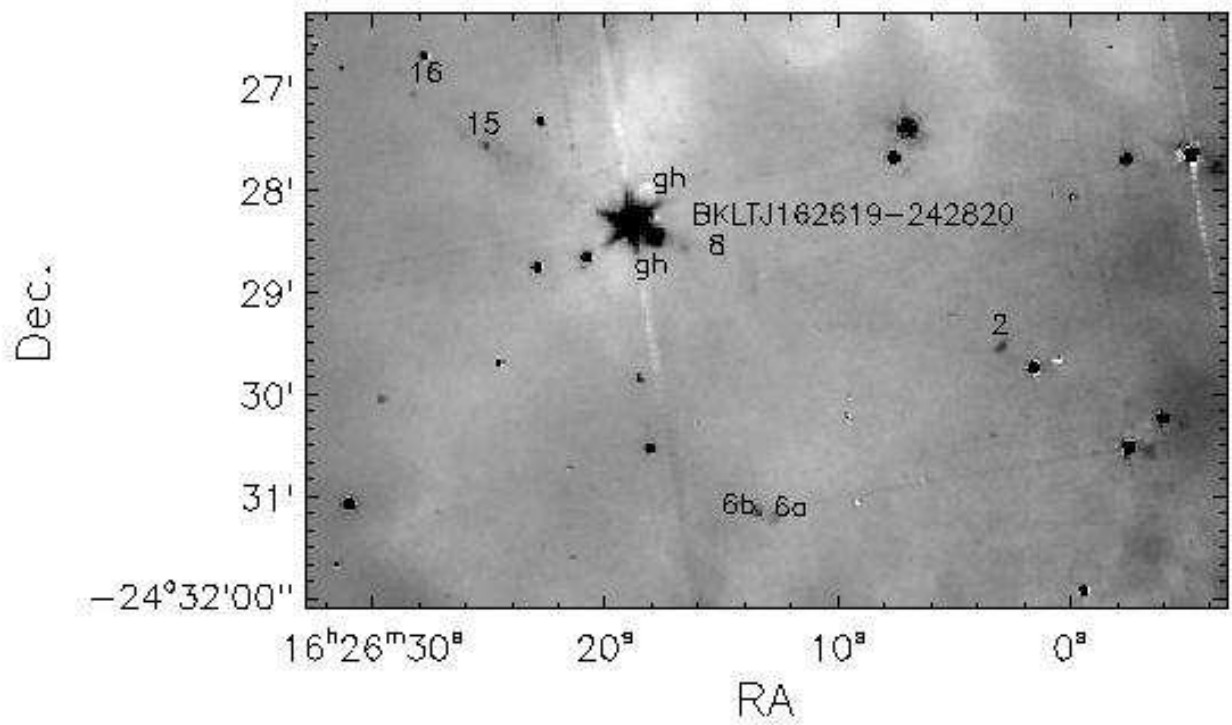}
\caption{The same as Fig.~\ref{fig3}, but for the EGOs 02, 06, 08, 15, and 16 region. The BKLT source is from \citet{bar97}.\label{fig20}}
\end{figure}

\begin{figure}
\epsscale{1.0}
\plotone{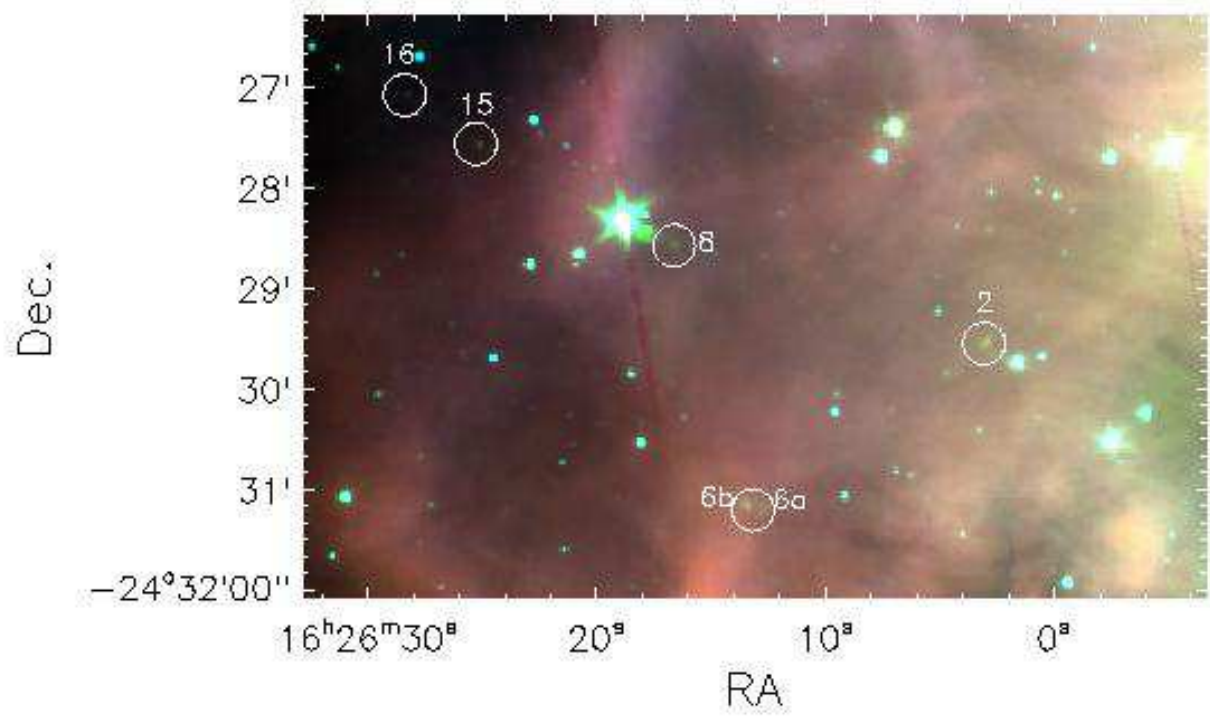}
\caption{The same as Fig.~\ref{fig4}, but for the EGOs 02, 06, 08, 15, and 16 region. \label{fig21}}
\end{figure}

EGO 41 is located about 5\farcm2 to the southwest of the Class 0 object IRAS 16293-2422  \citep{eva09} which drives two molecular outflows, one in the east-west direction and another one in the NE-SW direction \citep{woo87,cas01,hir01,lis02,sta04}. The IRAS 16293-2422 E-W molecular outflow also has its spectacular demonstration in the mid-infrared which is designated as EGO 43 and, together with EGOs 41 and 42, is shown in Fig.~\ref{fig13}. The scale of the EGO 43 outflow measures about 11\arcmin. It consists of  four main portions, EGO 43 A-D. The details of the inner portions A-B and the outer portions C-D of this outflow are shown in Figs.~\ref{fig14} and \ref{fig15}, respectively. IRAS 16293-2422 is a binary with components A and B \citep{mun92} and their positions are marked in Fig.~\ref{fig14} with pentagrams. IRAS 16293-2422 is visible in the MIPS 24 \micron \ image and invisible in the IRAC images. In Fig.~\ref{fig14}, the position of the submillimeter source IRAS 16293E \citep{nwa06}, 90\arcsec \ to the southeast of IRAS 16293-2422, is also marked.

From Fig.~\ref{fig13} we can see that the symmetry axis of the EGO 41 bowlike structure goes through EGO 42 and IRAS 16293-2422. This axis coincides well with the axis of the IRAS 16293-2422 NE-SW molecular outflow (see Fig.~12 in \citealt{lis02} ), suggesting that EGOs 41 and 42 may have physical association with this molecular outflow which is attributed to the A component of IRAS 16293-2422 \citep{sta04}.

We note that the EGO 41 bow shock points to source IRAS 16293-2422. This morphology usually suggests that the driving source is located on the direction opposite to IRAS 16293-2422. However, bow shocks are not always in a configuration of facing away from their driving sources. For example, the HH 311 bow shock \citep{rei97} and the southeastern bow shock of the V380 Ori-NE flow \citep{dav00} are in a configuration facing toward their exciting sources. This configuration can be understood if the flow impacts a dense, ambient clump, around which shocked gas is streaming \citep{dav00}. So it is also possible that EGO 41 is associated with IRAS 16293-2422.

EGO 44 is a bow shock in the IRAC channel 2 minus channel 1 image (Fig.~\ref{fig16}). It is visible in all the IRAC channels and Fig.~\ref{fig17} presents the three-color image of the region. EGO 44 is located at the east outskirts of the $\rho$ Ophiuchi cloud (Fig.~\ref{fig2}). The nearest known young stellar object to EGO 44 is IRAS 16367-2356 (Class II in \citealt{eva09}) which is about 15\arcmin \ to the west of EGO 44. As the EGO 44 bow shock faces to the east, IRAS 16367-2356 is likely the exciting source of EGO 44.

\subsection{IRAC Counterparts of Known Outflows}

Extensive surveys of outflows toward the $\rho$ Ophiuchi cloud, including HH objects in the optical and H$_2$ 2.12 \micron \ emission in the near-infrared, have been performed. In total, 46 HH objects, including components of HH objects, and 119 H$_{2}$ emission features have be detected in this region \citep{gro01,wu02,gom03,kha04,phe04}. We identified in the IRAC images 31 EGOs that correspond to the known outflows, among which seven EGOs correspond to known HH objects and 30 EGOs to H$_{2}$ near-infrared emission.

Figs.~\ref{fig18} and \ref{fig19} show the region of EGO 01 which has been identified at 2.12 \micron \ by \citet{luc08}. EGO 01 are three diffuse nebulae in the 2.12 \micron \ image while it consists of 3 faint knots in the 3.6 \micron \ and 4.5 \micron \ images. Two YSOs, GSS 23 and SR 4, are located in the nearby of EGO 01. GSS 23 is classified as a weak-line T Tauri star \citep{bou92}. \citet{bar02} and \citet{bit08} detected narrow H$_{2}$ emission surrounding GSS 23. SR 4 is an emission-line star \citep{str49} and \citet{phe04} detected an HH object, HH 312, to the southeast of SR 4. \citet{luc08} suggested that there are some connection between the H$_{2}$ emission and the outflow driving by VLA1623-243 (see Figs.~\ref{fig24}-~\ref{fig26} and discussions on EGOs in that region) on the basis of their locations and morphology.

Figs. 5-6 have shown the region of EGO 34 which are counterparts of the outflow [KGS2004] f05-04 \citep{kha04}. \citet{kha04} and \citet{smi05} observed three knots which correspond to EGO 34a-c in their near-infrared observations. However the {\it Spitzer} images show much more details of the outflow than the near-infrared images. In Figs.~\ref{fig5} and \ref{fig6} we can see extended diffuse emission to the northeast of EGO 34c and to the southwest of EGO 34a. The central source of the outflow, MMS126, which is identified as a low-mass Class 0 object in the millimetre continuum observations by \citet{sta06} (Classified as Class 0 in \citealt{eva09}) is visible in all IRAC bands and also in the MIPS 24 \micron \ image. \citet{sta06} has observed a molecular CO outflow from MMS126 which is orientated in the NE-SW direction.

Figs.~\ref{fig20} and \ref{fig21} present the images of EGOs 02, 06, 08, 15, and 16 that corresponds to near-infrared flows [KGS2004] f10-01 and [KGS2004] 10-02 \citep{kha04}. \citet{kha04} detected nine H$_{2}$ emission components of [KGS2004] f10-01. However, only four knots, EGOs 02, 08, 15, and 16, can be detected in the IRAC images. These four EGOs are located on the two sides of BKLT J162619-242820, a Class II object \citep{eva09}. \citet{kha04} proposed BKLT J162619-242820 as the driving source of this flow. EGO 06 has two components, EGO 06a and 06b, corresponding to [KGS2004] f10-02. The driving source of EGO 06 is unclear.

\begin{figure}
\epsscale{1.0}
\plotone{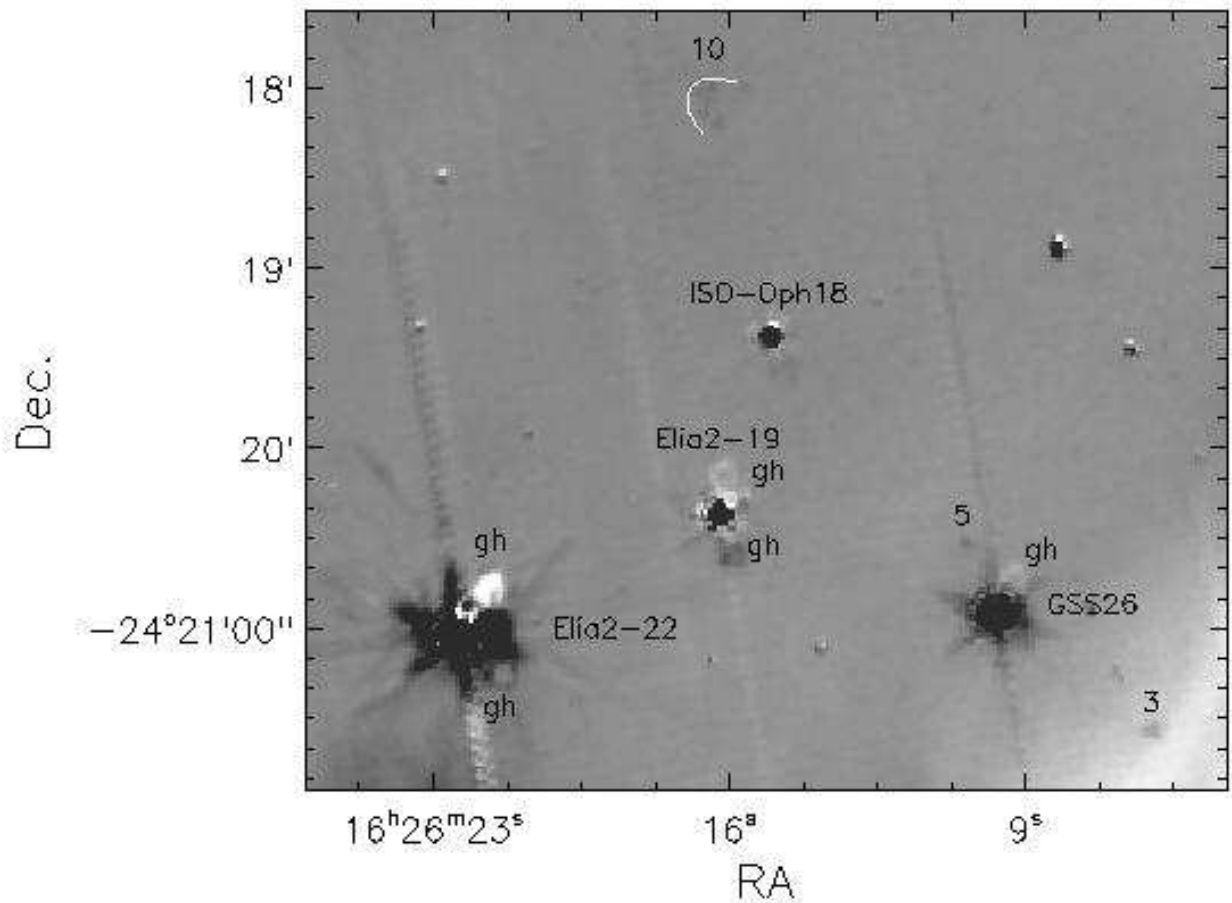}
\caption{The same as Fig.~\ref{fig3}, but for the region of EGOs 03, 05, and 10. Elias 2 is from \citet{eli78}, GSS 26 from \citet{gra73}, and ISO-oph 18 from \citet{bon01}. The bow shock morphology of EGO 10 is outlined with a white line. \label{fig22}}
\end{figure}

\begin{figure}
\epsscale{1.0}
\plotone{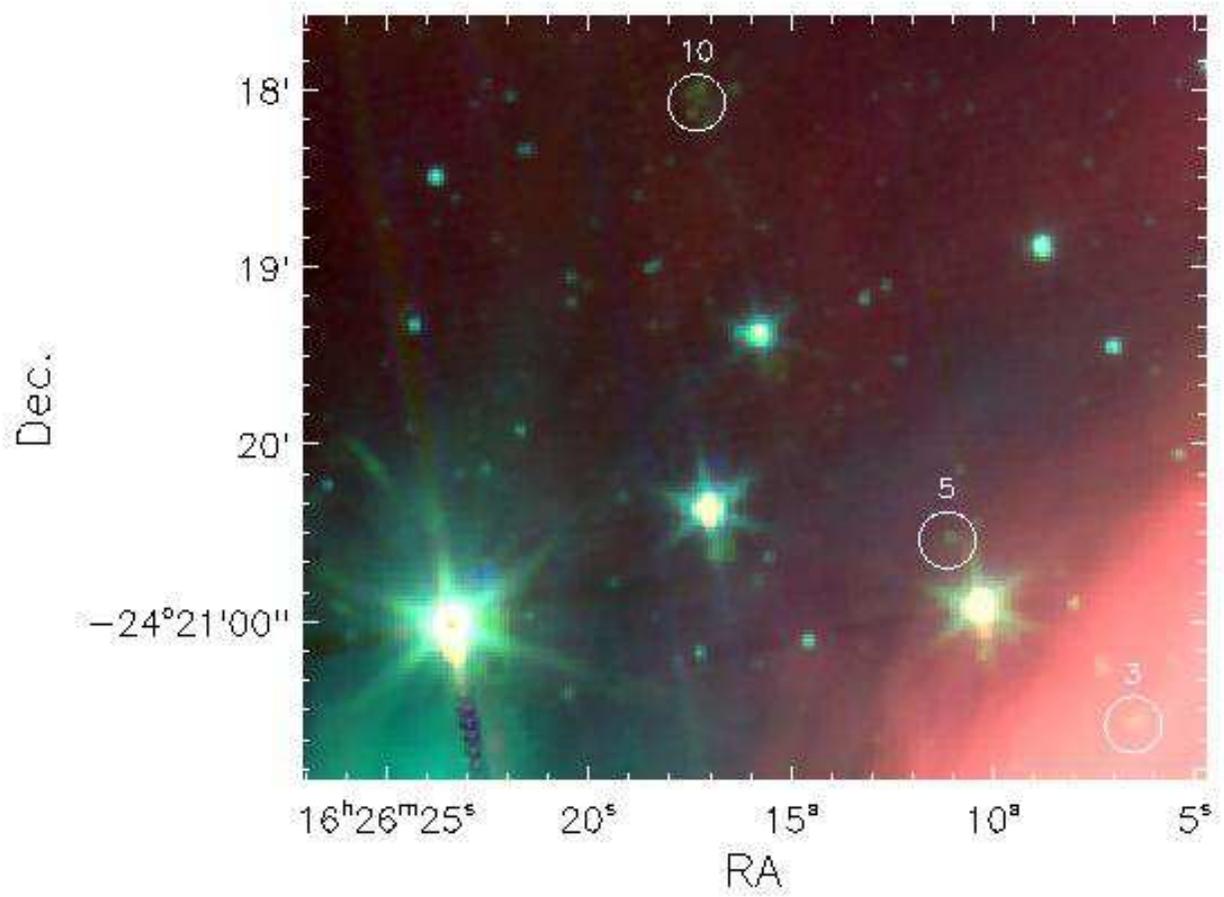}
\caption{The same as Fig.~\ref{fig4}, but for the region of EGOs 03, 05, and 10. \label{fig23}}
\end{figure}

\begin{figure}
\epsscale{1.0}
\plotone{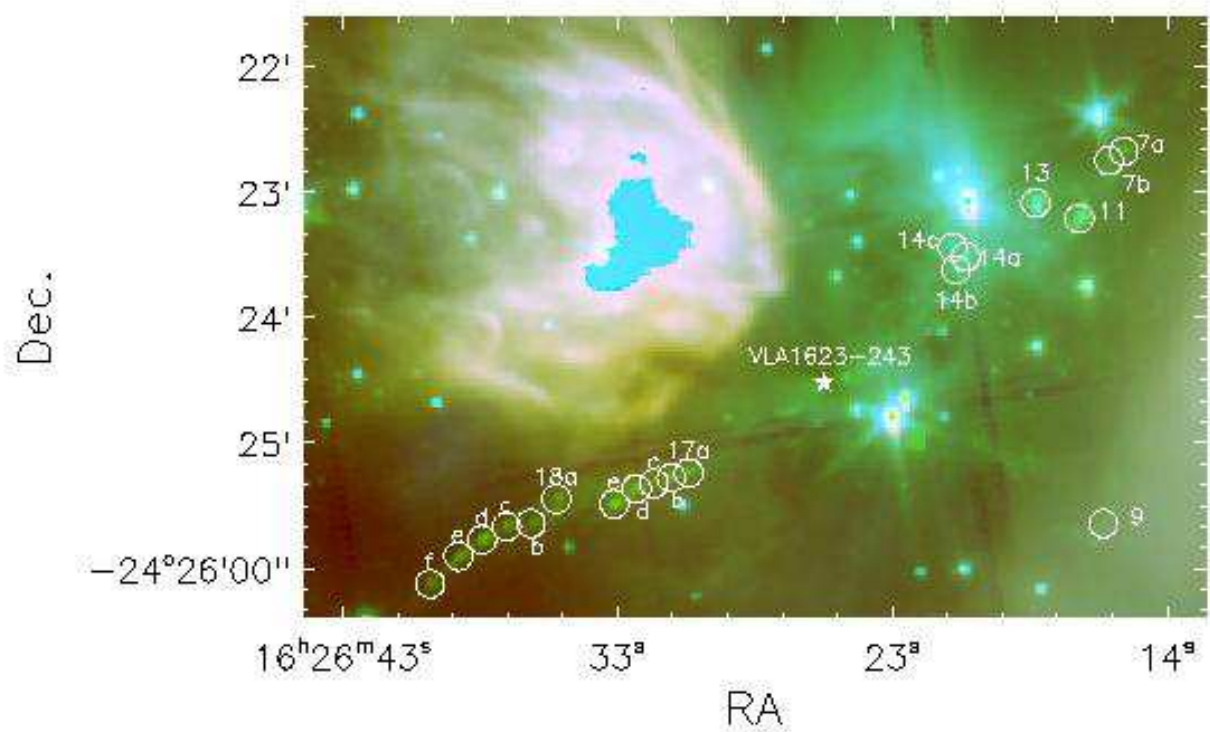}
\caption{The same as Fig.~\ref{fig4}, but for the region of EGOs 07, 09, 11, 13-14, 17-18. The location of the Class 0 sources VLA1623-243 is marked with a white pentagram. \label{fig24}}
\end{figure}

\begin{figure}
\epsscale{1.0}
\plotone{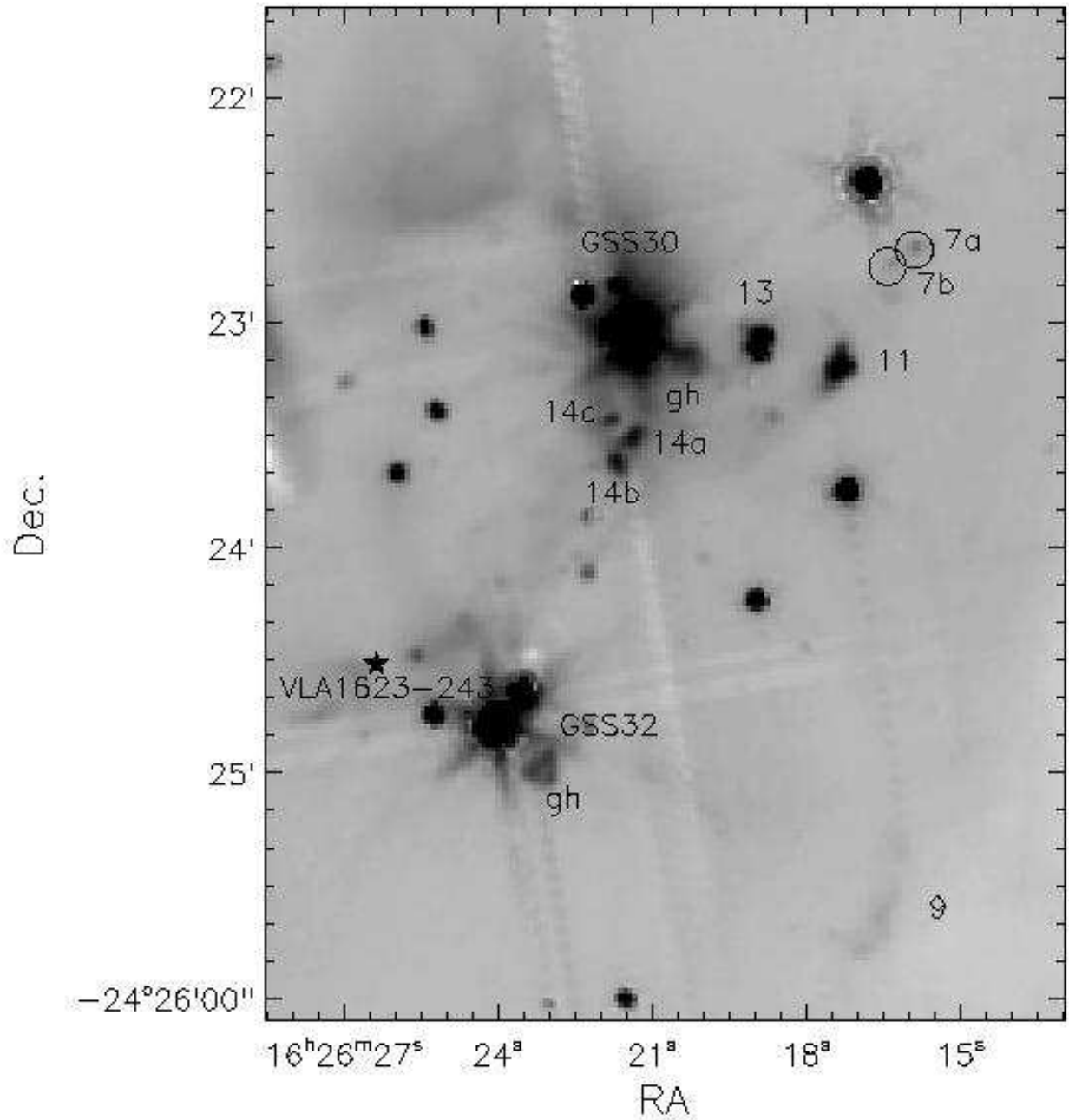}
\caption{The same as Fig.~\ref{fig3}, but for the VLA1623-243 northwest region. The location of the Class 0 sources VLA1623-243 is marked with a pentagram. The GSS source is from \citet{gra73}. \label{fig25}}
\end{figure}

\begin{figure}
\epsscale{1.0}
\plotone{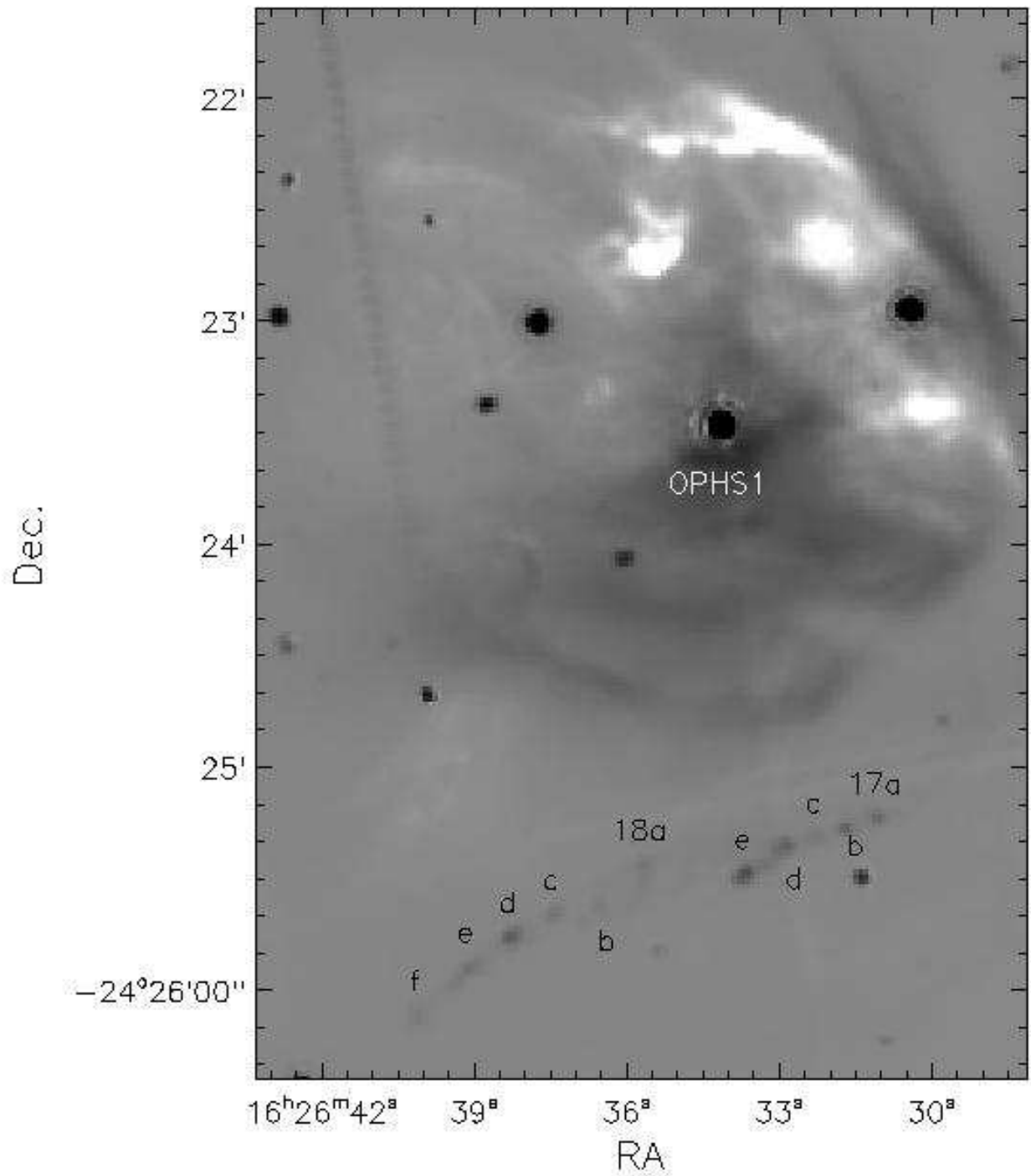}
\caption{The same as Fig.~\ref{fig3}, but for the VLA1623-243 southeast region. The OPH source is from \citet{gra73}. \label{fig26}}
\end{figure}

The IRAC images of EGOs 03, 05, and 10 are presented in Figs.~\ref{fig22} and \ref{fig23}. EGO 03 is the counterpart of [GSWC2003] 8a and EGO 05 is the counterpart of [GSWC2003] 8d. EGO 10 corresponds to [GSWC2003] 9 \citep{gom03}. \citet{gom03} identified [GSWC2003] 8a as three distinct components. In Figs.~\ref{fig22} and \ref{fig23} EGO 03 is a diffuse patch. GSS 26, a Class II source \citep{eva09}, is the nearest YSO to EGO 03 and EGO 05 and may be the driving source for these EGOs. EGO 10 is a diffuse structure within which we can identify several relatively bright knots. The components of EGO 10 constitute a faint bow-like structure with its symmetry axis pointing to GSS 26, suggesting that EGO 10 is also driven by this source.

Fig.~\ref{fig24} shows the locations of EGOs 07, 09, 11, 13-14, and 17-18. Their previous identifications in optical and near-infrared can be found in Table~\ref{tab3}. \citet{gom03} observed this region and proposed that most of the emission features in this region are driven by VLA1623-243, a Class 0 source in L1688 \citep{and93} which is marked with a pentagram in Fig.~\ref{fig24}. VLA1623-243 has been identified as the driving source of a large-scale molecular outflow which lies in the direction of NW-SE \citep{and90}. \citet{den95} presented maps of the large-scale $^{12}$CO outflow and H$_2$ near-infrared emission images of this region.

Fig.~\ref{fig25} shows the northwest part of the VLA1623-243 region. EGOs 07, 11, and 13-14 distribute roughly as a chain in the NW-SE direction. \citet{car06} observed this region in near-infrared and obtained proper motion measurements for the knots in the VLA1623 region. EGO 11 corresponds to HH 313B, which is likely associated with GSS 30, a class I binary system candidate in \citet{chen07}, as the position angle of proper motion measured for HH 313B is about 243\arcdeg \citep{car06}. EGO 09 is a diffuse nebula on the bottom right of Fig.~\ref{fig25}. \citet{kha04} detected this feature in the near-infrared and suggested that it is driven by a source to the north.

Fig.~\ref{fig26} shows the VLA1623-243 southeast region. We can see a chain of knots in the EGO 17 and EGO 18 flows. EGO 17 and EGO 18 correspond to near-infrared flows [GSWC2003] 20 and 21, respectively \citep{gom03}. The bright mid-infrared emission knots are visible at the IRAC 3.6 \micron, 4.5 \micron, and 5.8 \micron \ images while the faint ones are only visible in the 3.6 \micron \ and 4.5 \micron \ images.

\begin{figure}
\epsscale{1.0}
\plotone{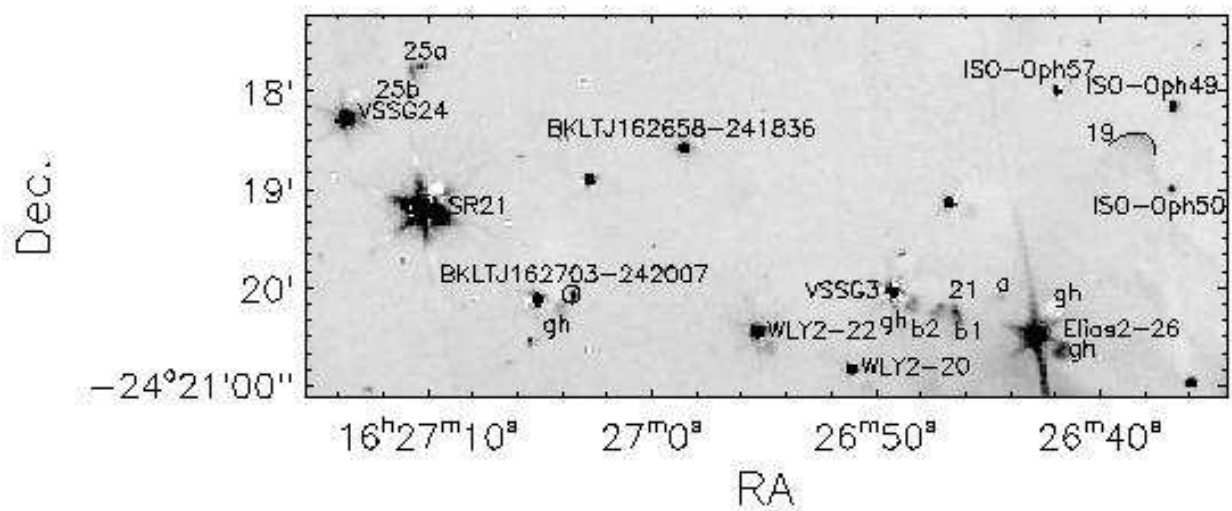}
\caption{The same as Fig.~\ref{fig3}, but for the EGOs 19, 21, and 25 region. Source designations used are as follows: BKLT sources are from \citet{bar97}, Elias sources from \citet{eli78}, SR sources from \citet{str49}, VSSG sources from \citet{vrb75} and WLY sources from \citet{wil89}. The bow shock morphology of EGO 19 is outlined with a black line.\label{fig27}}
\end{figure}

\begin{figure}
\epsscale{1.0}
\plotone{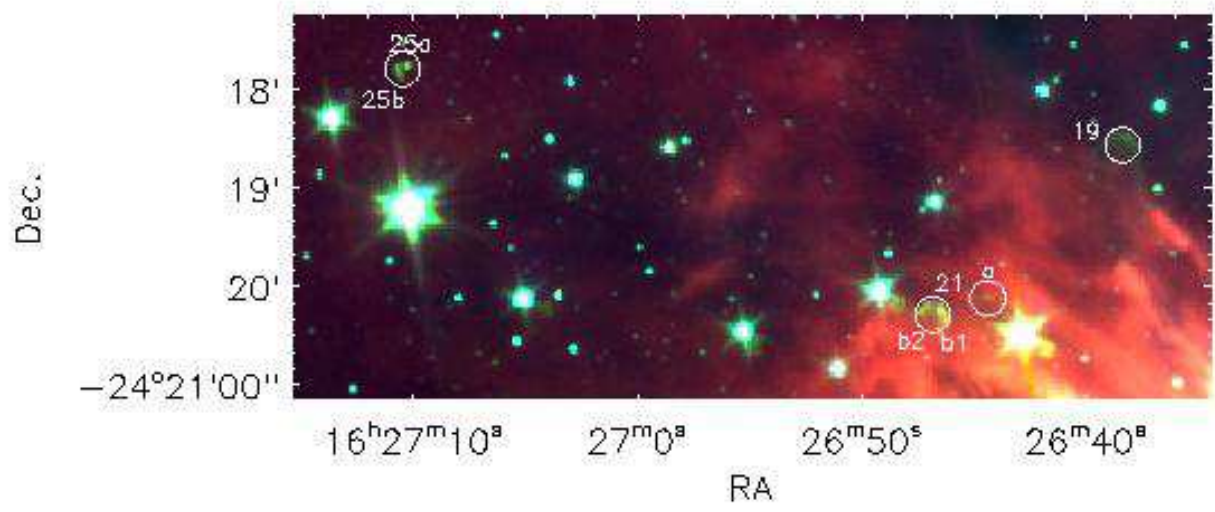}
\caption{The same as Fig.~\ref{fig4}, but for the EGOs 19, 21, and 25 region. \label{fig28}}
\end{figure}

The IRAC images of EGOs 19, 21, and 25 are shown in Figs.~\ref{fig27} and \ref{fig28}. EGO 19 is a bow-like structure which corresponds to [GSWC2003] 24a in the near-infrared \citep{gom03} and an HH object candidate in the optical \citep{phe04}. \citet{gom03} suggested Elias 2-26 (Class II in \citealt{eva09}) as the driving source of [GSWC2003] 24a. The distance from Elias 2-26 to EGO 19 is about 2\farcm2. EGO 21a is a faint patch and EGOs 21b1 and 21b2 are elongated knots. EGOs 21b1 and 21b2 correspond to HH 79a1 and a2 in the optical image \citep{phe04}. The relationship between EGO 21a and EGO 21b is uncertain and they could be excited by different exciting sources \citep{gom03,phe04}. EGO 25 consists of two bright knots on the top left of Figs.~\ref{fig27} and \ref{fig28}. They correspond to [GSWC2003] 25a and 25b in the near-infrared images \citep{gom03} and HH 711 in the optical \citep{phe04}. \citet{gom03} suggested three possible driving sources for this outflow, including BKLT J162658-241836, VSSG 24, and SR 21. In contrast, \citet{phe04} proposed that VSSG 3 is the exciting source of HH 711.

\begin{figure}
\epsscale{1.0}
\plotone{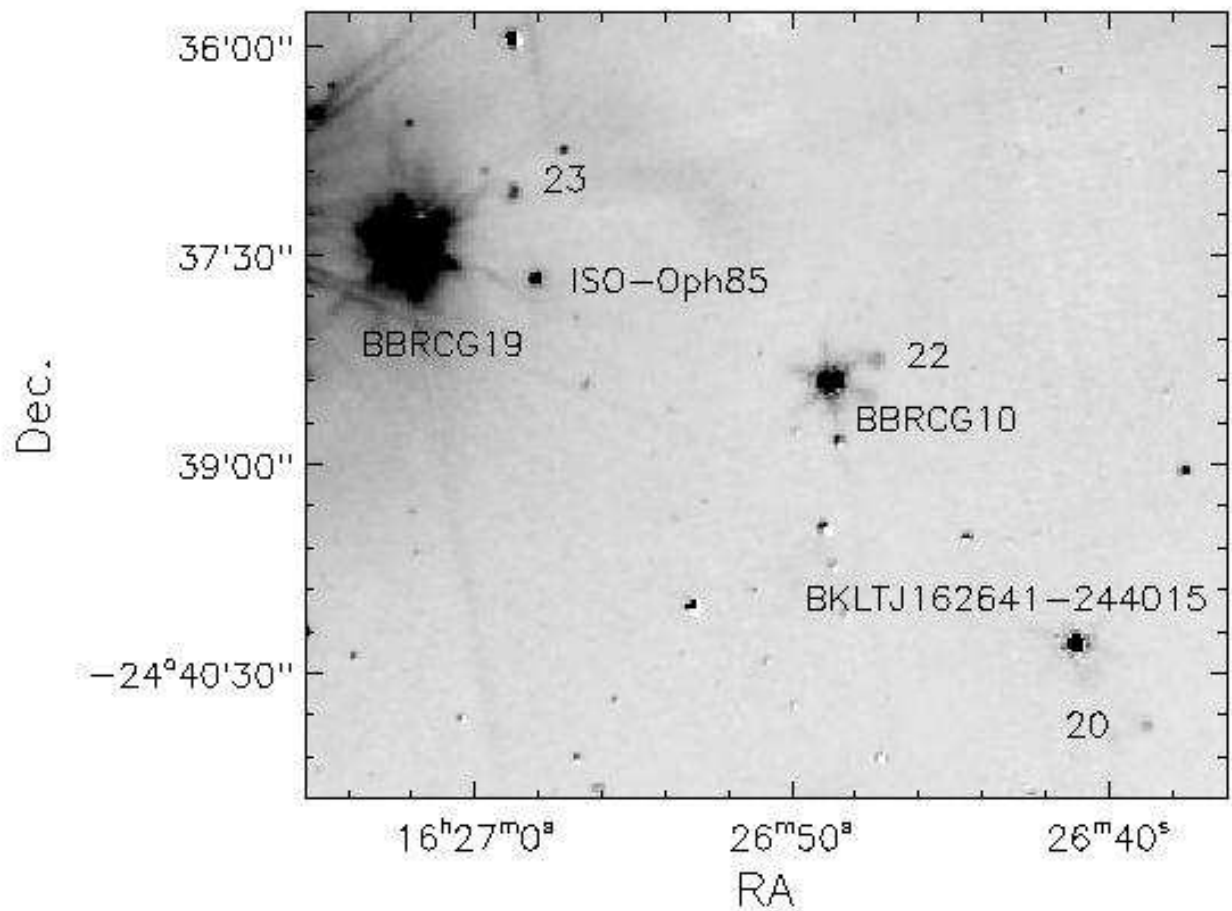}
\caption{The same as Fig.~\ref{fig3}, but for the EGOs 20 and 22-23 region. Source designations used are as follows: BKLT sources are from \citet{bar97}, ISO-oph sources from \citet{bon01} and BBRCG sources from \citet{bar89}. \label{fig29}}
\end{figure}

\begin{figure}
\epsscale{1.0}
\plotone{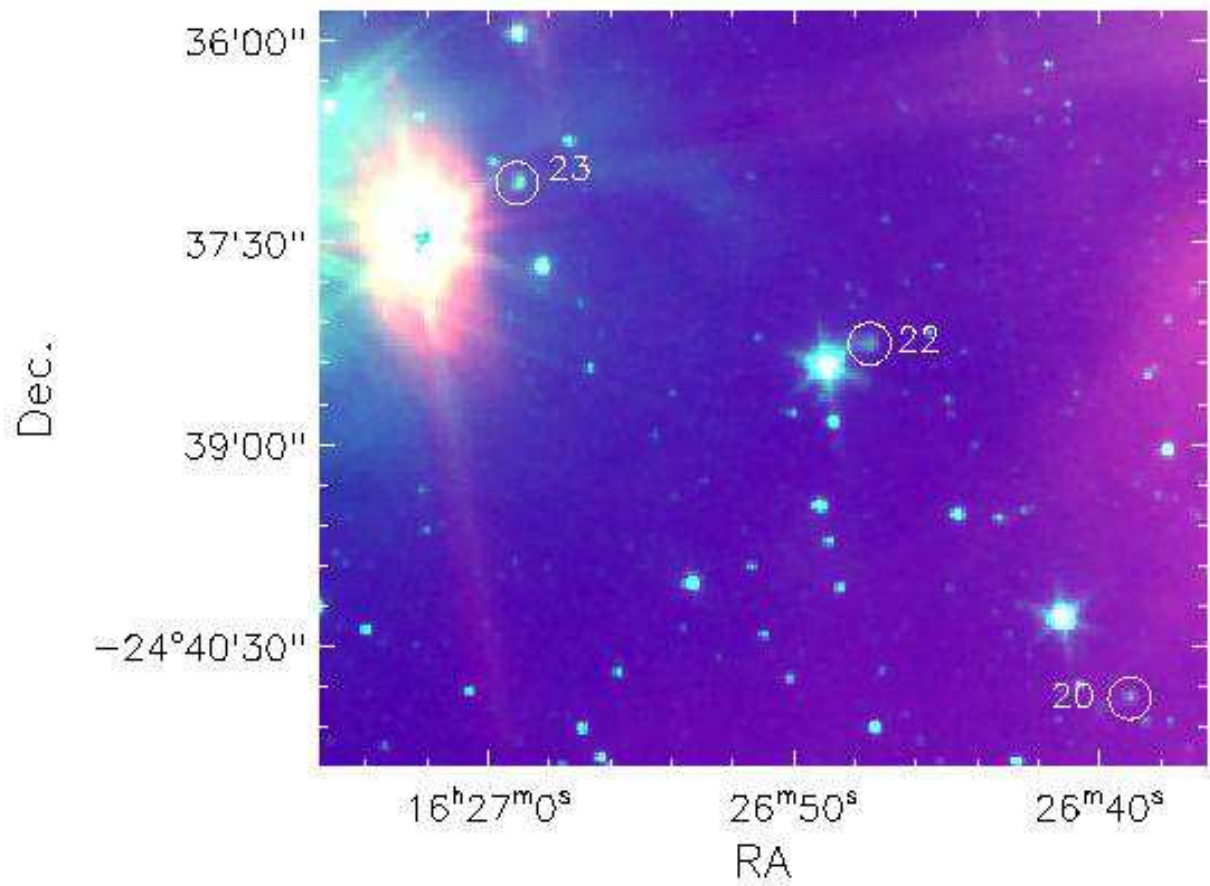}
\caption{The same as Fig.~\ref{fig4}, but for the EGOs 20 and 22-23 region. \label{fig30}}
\end{figure}

\begin{figure}
\epsscale{1.0}
\plotone{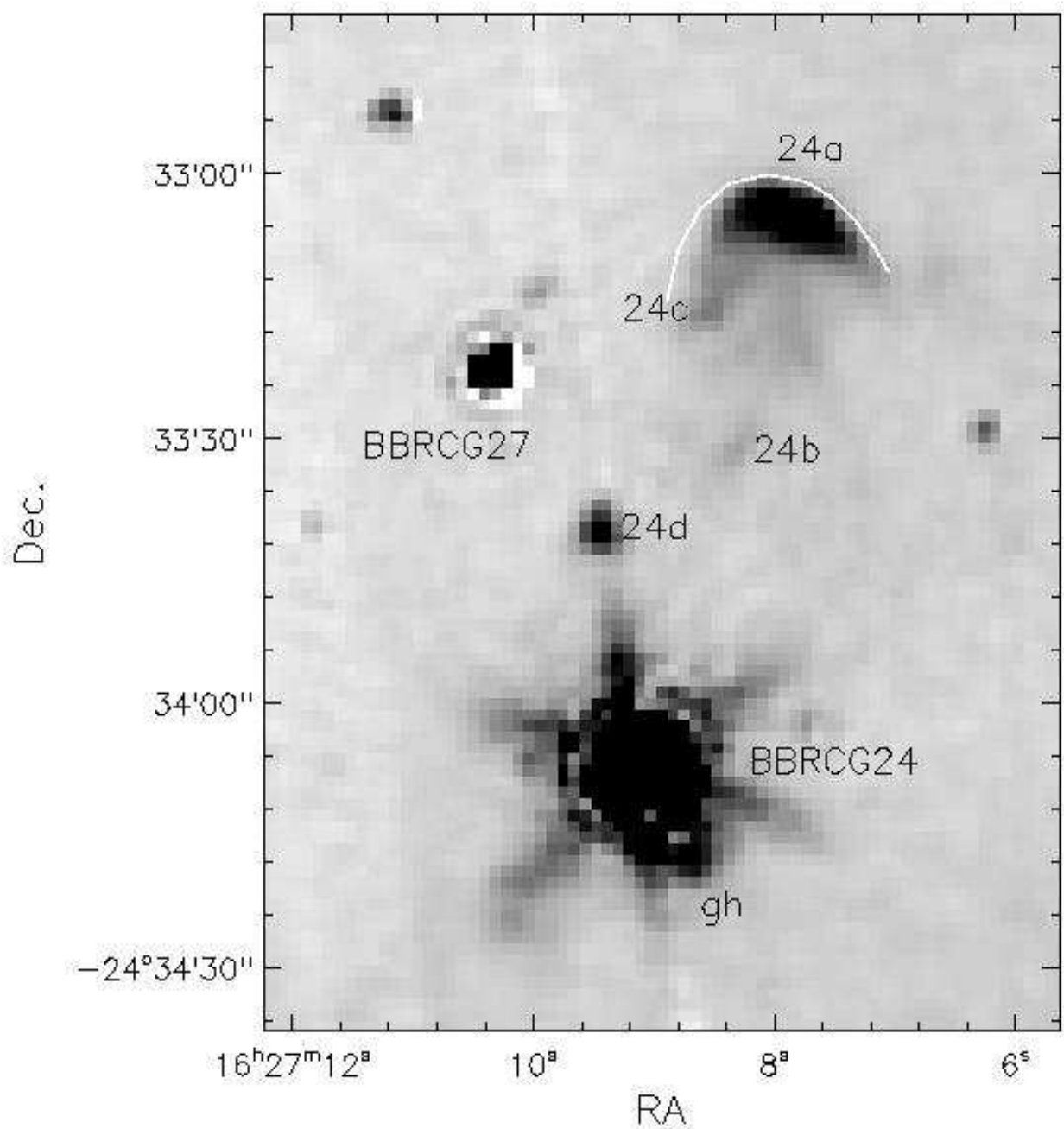}
\caption{The same as Fig.~\ref{fig3}, but for the EGO 24 region. BBRCG sources are from \citet{bar89}. The bow shock morphology of EGO 24a is outlined with a white line.\label{fig31}}
\end{figure}

\begin{figure}
\epsscale{1.0}
\plotone{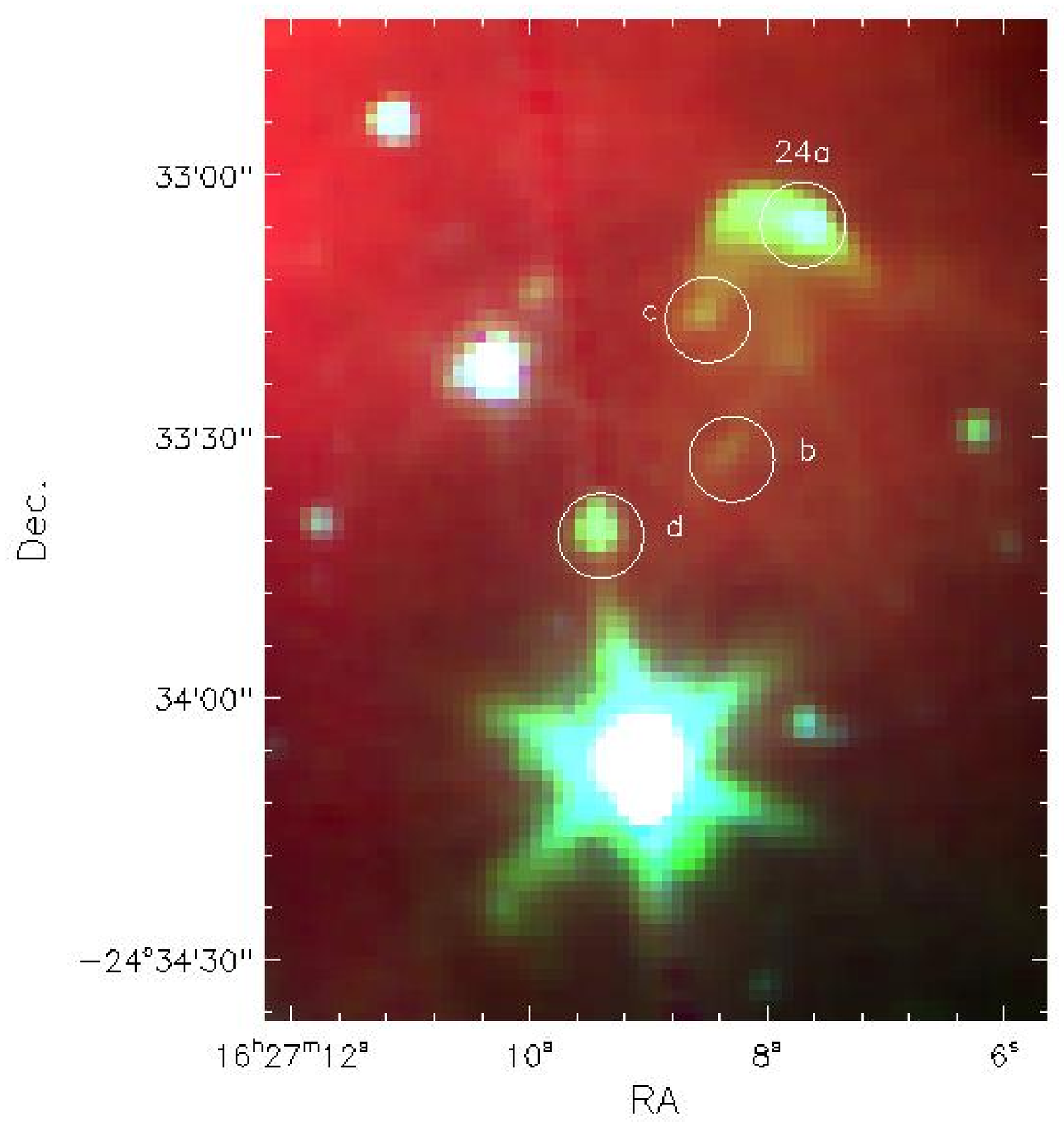}
\caption{The same as Fig.~\ref{fig4}, but for the EGO 24 region.\label{fig32}}
\end{figure}

\begin{figure}
\epsscale{1.0}
\plotone{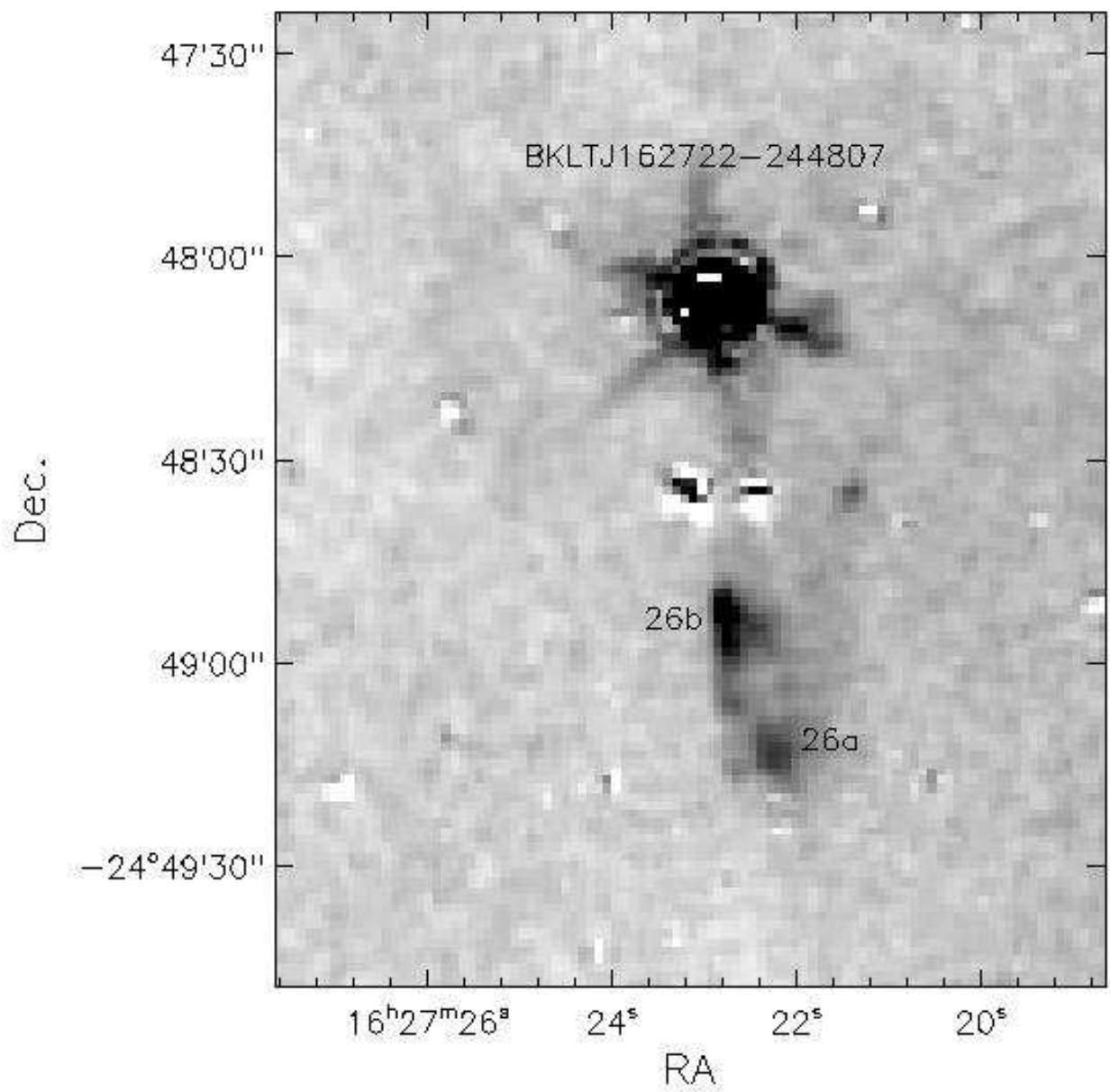}
\caption{The same as Fig.~\ref{fig3}, but for the EGO 26 region. The BKLT source is from \citet{bar97}.\label{fig33}}
\end{figure}

\begin{figure}
\epsscale{1.0}
\plotone{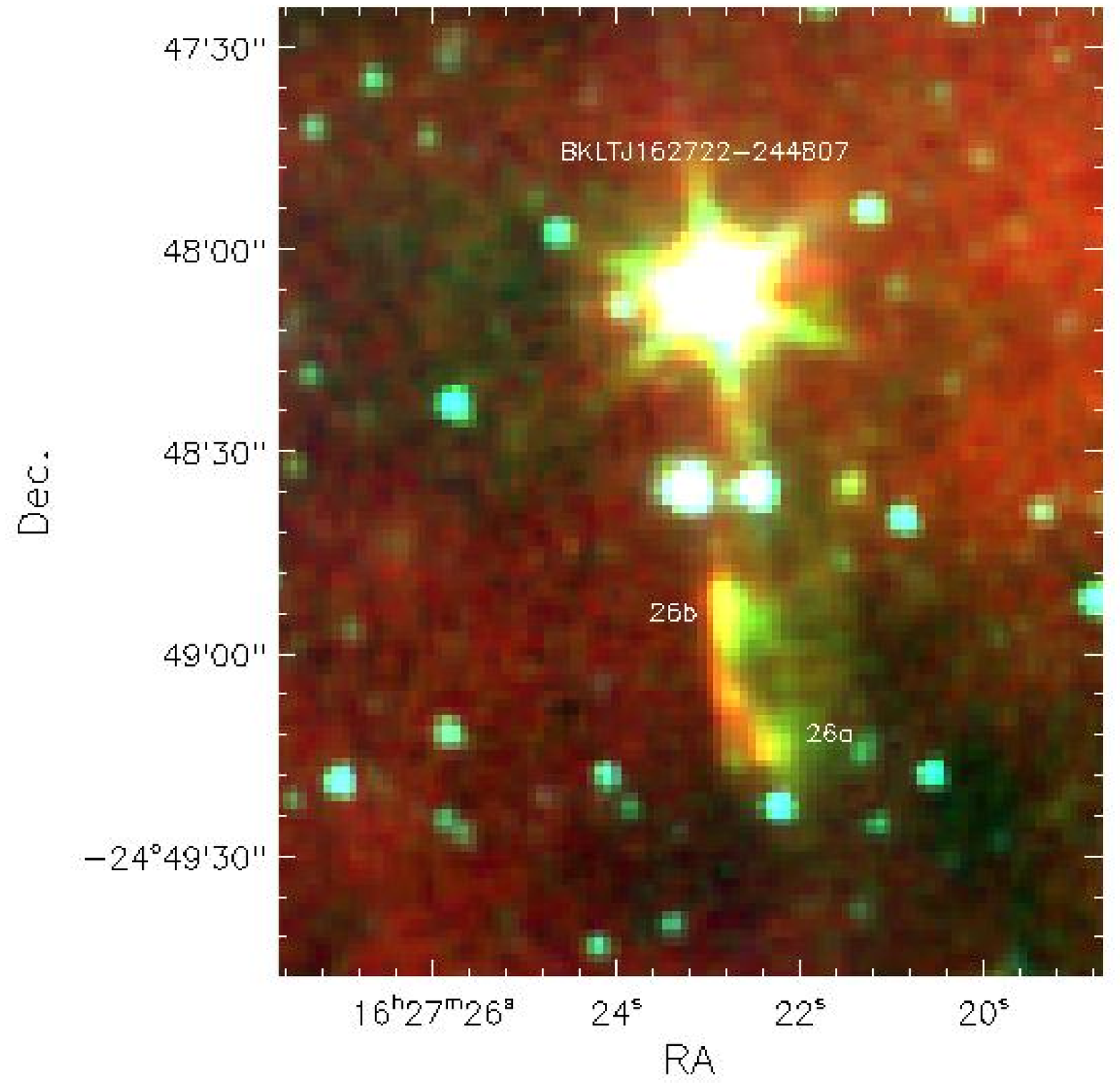}
\caption{The same as Fig.~\ref{fig4}, but for the EGO 26 region. \label{fig34}}
\end{figure}

Figs.~\ref{fig29} and \ref{fig30} show the region of EGOs 20 and 22-23. EGO 20 is a faint knot on the bottom right of Figs.~\ref{fig29} and \ref{fig30}. EGO 20 corresponds to a known HH object, HH 314 \citep{phe04}. EGO 22 is a diffuse knot which corresponds to [KGS2004] f04-01 in the near-infrared and HH 673 \citep{phe04} in the optical. EGO 23 has a bow shock shape and coincides to [KGS2004] f04-01b. There are three YSOs in this region, BKLT J162641-244015 (Class II in \citealt{eva09}), BBRCG10 (Class II in \citealt{eva09}), and BBRCG19 (Class I, \citealt{wil89}). \citet{phe04} proposed BKLT J162641-244015 as the driving source of HH 314.

Figs.~\ref{fig31} and \ref{fig32} show the region of EGO 24, which consists of components EGOs 24a-d. These components correspond to four emission knots in the near-infrared flow [KGS2004] f09-01. In this region \citet{kha04} detected seven H$_2$ emission knots, f09-01a-g. EGO 24a has a bow shock shape and EGO 24d is a bright knot while EGO 24b-c are faint diffuse nebulae. EGO 24a and EGO 24c may belong to the same bow shock structure. \citet{gom03} suggested that BBRCG 24 (Class II in \citealt{eva09}) is the driving source of the emission feature corresponding to EGO 24d and that WL 16, a early type embedded star \citep{pad08} located about 3\arcmin \ south of EGO 24a, might be responsible for the emission features corresponding to EGO 24a-c. \citet{kha04} suggested another possibility that f09-01a-g belong to the same flow.

\begin{figure}
\epsscale{1.0}
\plotone{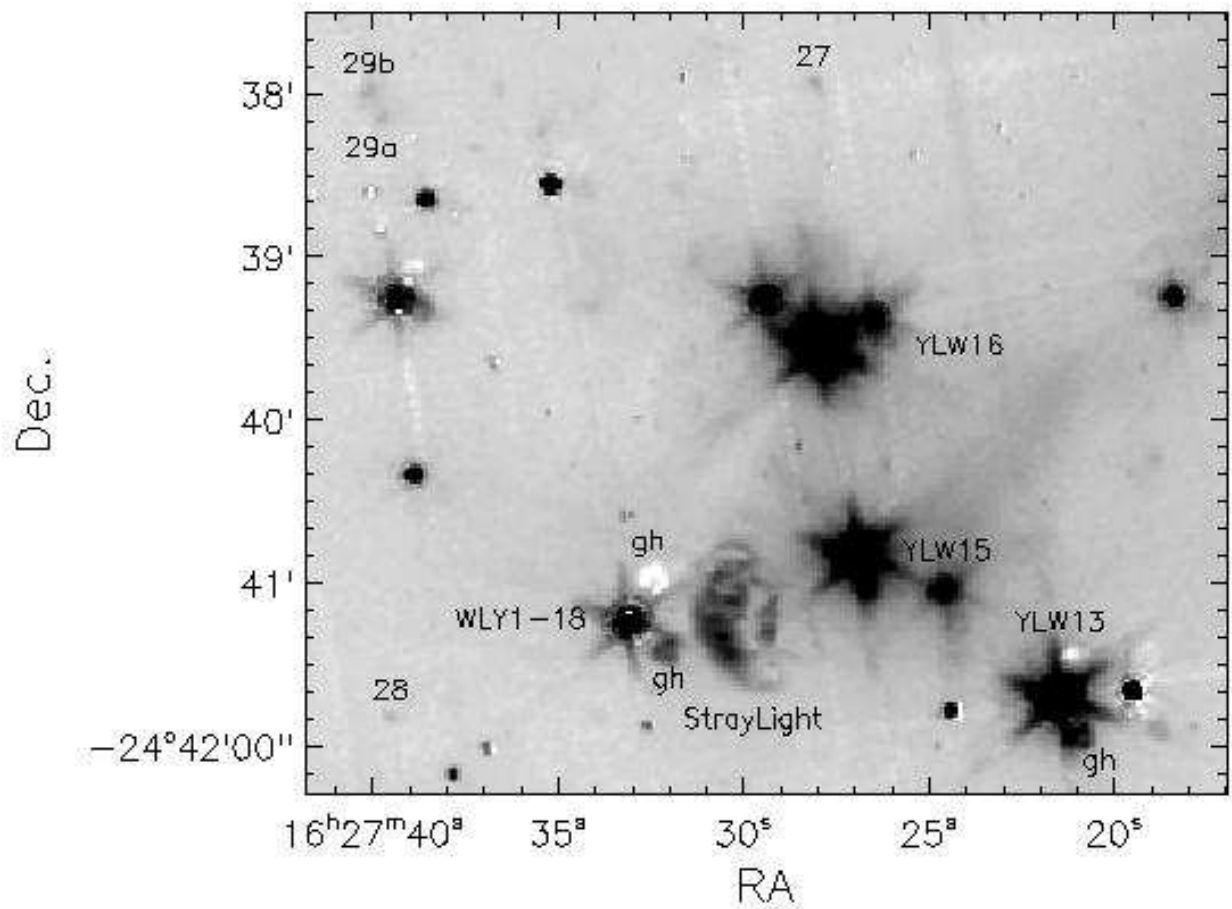}
\caption{The same as Fig.~\ref{fig3}, but for the EGOs 27-29 region. YLW sources are from \citet{you86} and WLY1 sources from \citet{wil89}. \label{fig35}}
\end{figure}

\begin{figure}
\epsscale{1.0}
\plotone{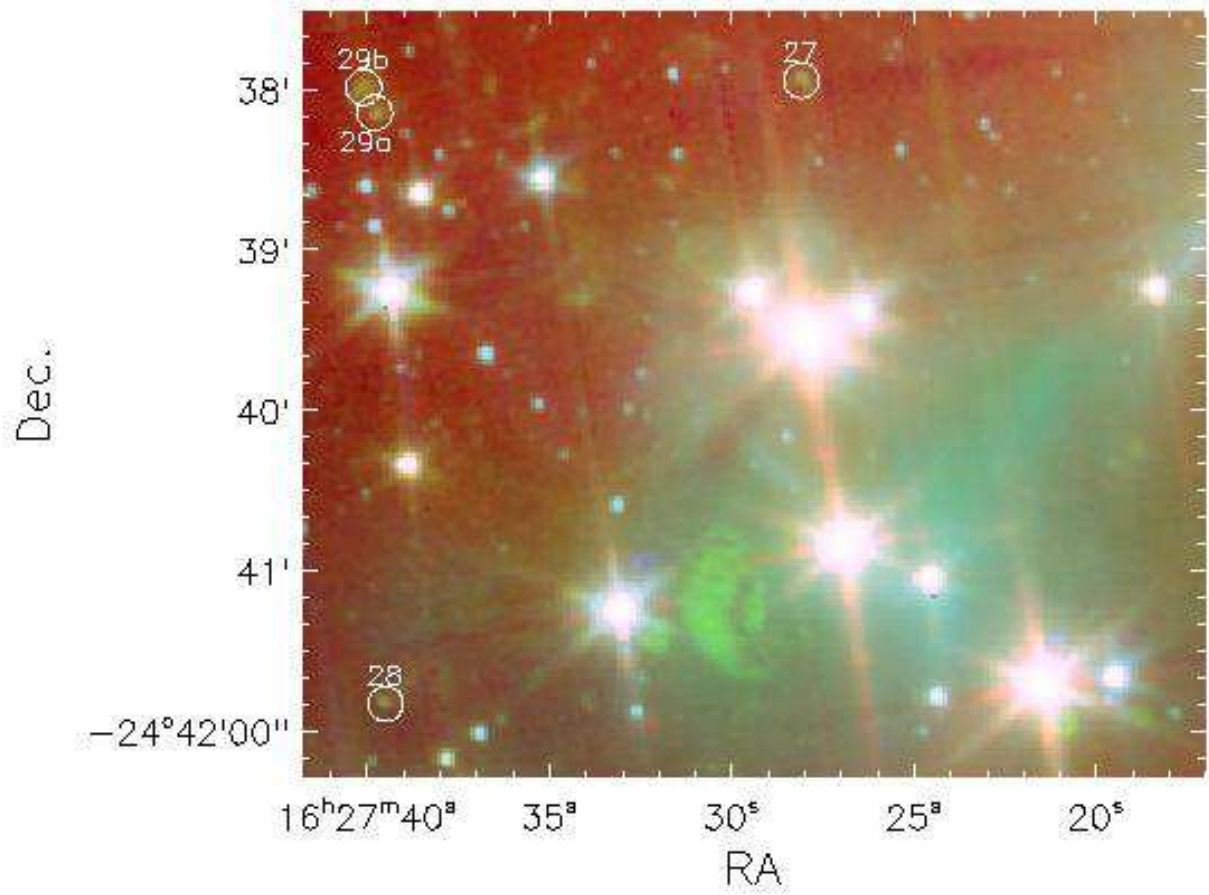}
\caption{The same as Fig.~\ref{fig4}, but for the EGOs 27-29 region. \label{fig36}}
\end{figure}

\begin{figure}
\epsscale{1.0}
\plotone{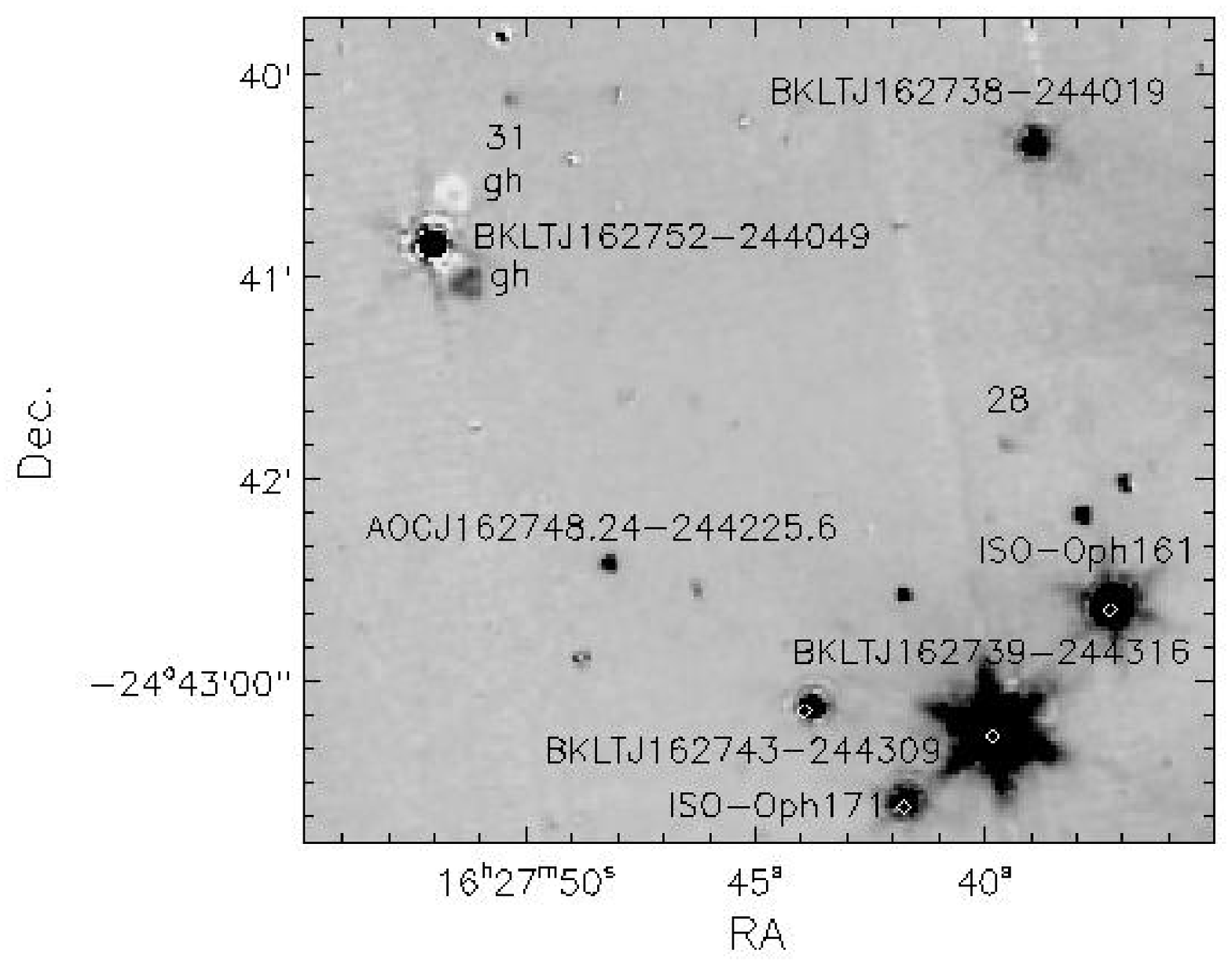}
\caption{The same as Fig.~\ref{fig3}, but for the EGOs 28 and 31 region. BKLT sources are from \citet{bar97}, ISO-oph sources from \citet{bon01}, and AOC sources from \citet{aoc08}. \label{fig37}}
\end{figure}

\begin{figure}
\epsscale{1.0}
\plotone{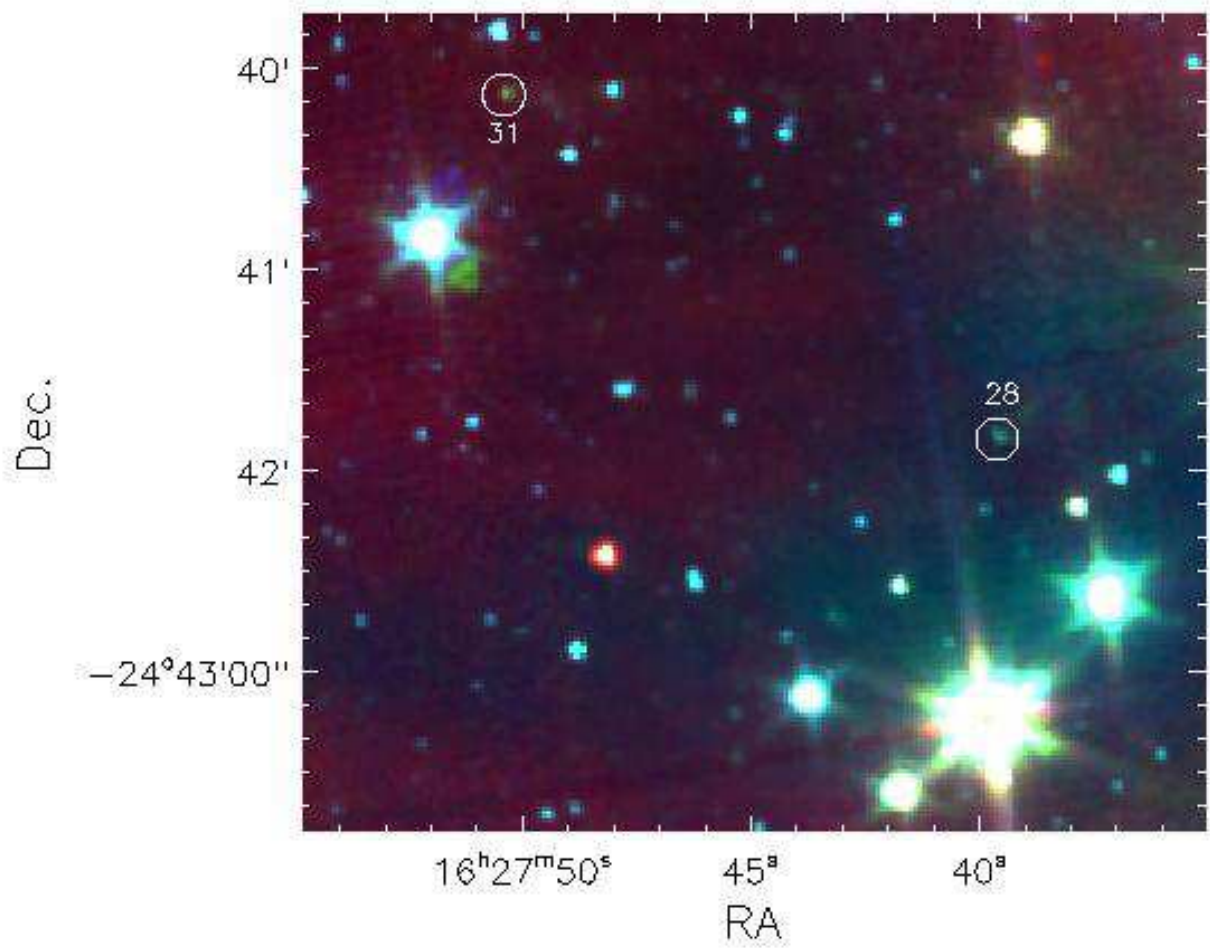}
\caption{The same as Fig.~\ref{fig4}, but for the EGOs 28 and 31 region. \label{fig38}}
\end{figure}

Figs.~\ref{fig33} and \ref{fig34} show the images of EGO 26 that coincides to [KGS2004] f03-02 \citep{kha04} in the near-infrared. EGO 26 also corresponds to an HH object, HH 224S \citep{phe04}. BKLT J162722-244807, a Class II source in \citet{eva09}, is the nearest protostar to this EGO. However, considering the association between [KGS2004] f03-02 and other flows in this region, neither \citet{kha04} nor \citet{phe04}  suggest this object as the exciting source of the emission feature corresponding to EGO 26. \citet{phe04} detected several [SII] emission features named HH 224N, HH 224NW1 and HH 224 NW2 to the northwest of HH 224S. They believed that HH 224S belongs to a large flow driven by an embedded source located to the northwest of EGO 26. \citet{phe04} suggested GY 193, a YSO located in the northwest about 7 \arcmin \ away from HH 224S, as the likely exciting source. \citet{kha04} detected the H$_{2}$ emission counterpart of HH 224S, but they did not detect the counterparts of HH 224N and HH 224NW. Therefore, they suggested EM*SR 24, which is about 8 \arcmin \ to the northwest of [KGS2004] f03-02, as the exciting source based on the fact that [KGS2004] f03-02, [KGS2004] f03-01, and EM*SR 24 are aligned along the northwest-southeast direction. In the IRAC images, only the counterpart of HH 224S, i.e., EGO 26, is detected. As EGO 26a and EGO 26b form a wide-cavity lobe which faces away from BKLT J162722-244807, we propose BKLT J162722-244807 to be the exciting source of EGO 26. Figs.~\ref{fig35} and \ref{fig36} show the region of EGOs 27-29, and EGO 31 is shown in Figs.~\ref{fig37} and \ref{fig38}. EGOs 27-29 all are faint knots. \citet{kha04} proposed YLW 15 (Class I, \citealt{wil89}) to be the likely driving source of the near-infrared emission feature corresponding to EGO 29. There are abundant YSOs in this region and the nearby YSOs are marked in Figs.~\ref{fig35} and ~\ref{fig37}.

\begin{figure}
\epsscale{1.0}
\plotone{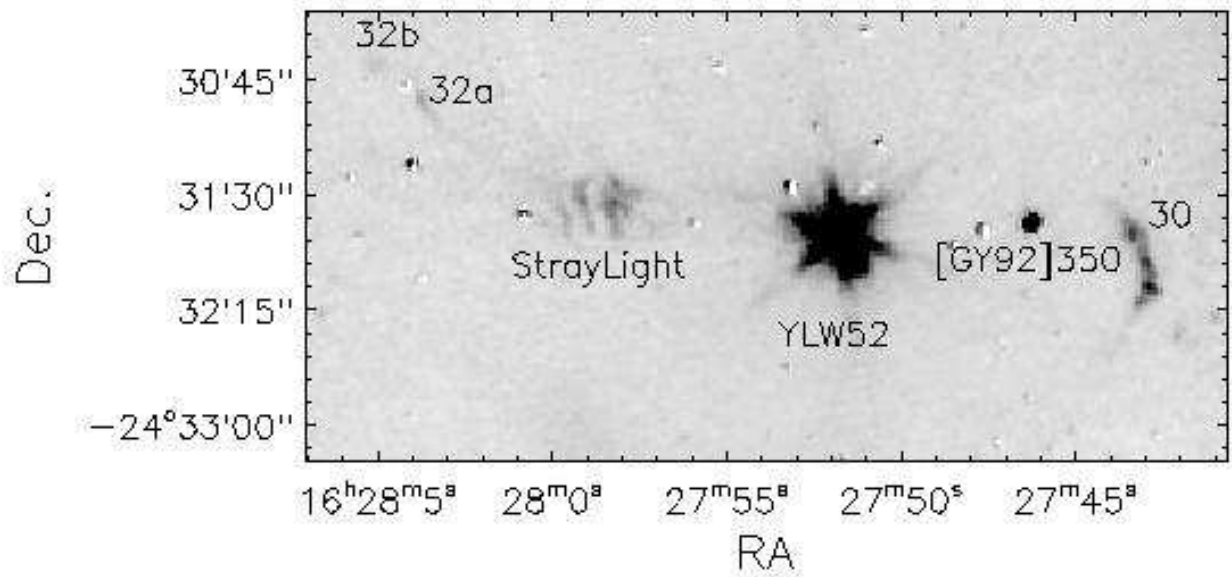}
\caption{The same as Fig.~\ref{fig3}, but for the EGOs 30 and 32 region. The YLW source is from \citet{you86} and the GY92 source is from \citet{gre92}. \label{fig39}}
\end{figure}

\begin{figure}
\epsscale{1.0}
\plotone{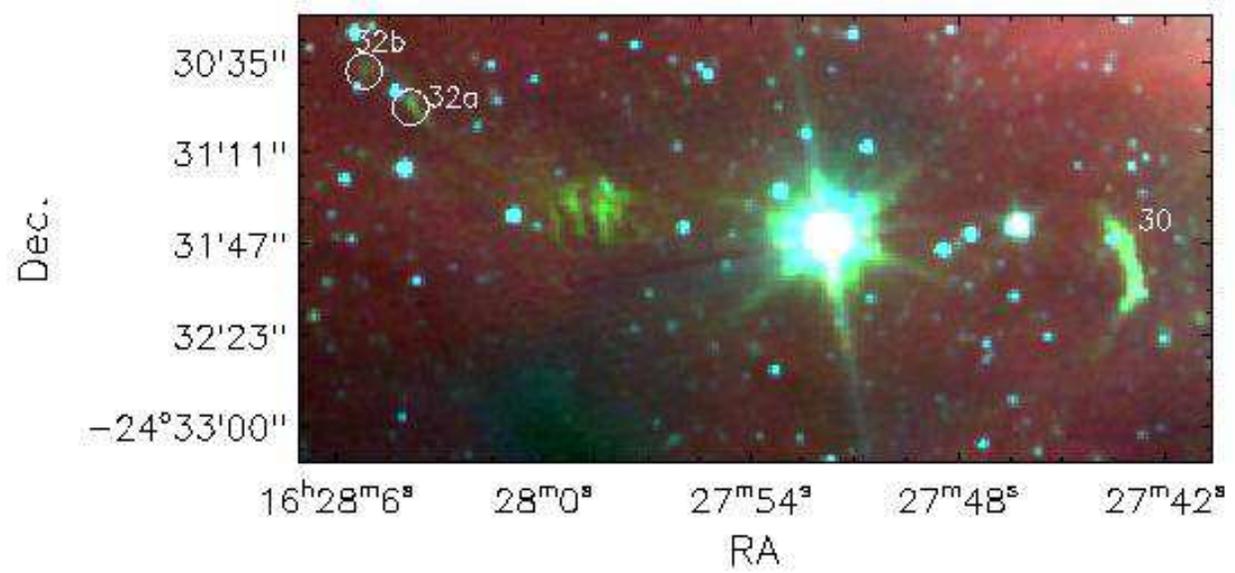}
\caption{The same as Fig.~\ref{fig4}, but for the EGOs 30 and 32 region. \label{fig40}}
\end{figure}

Figs.~\ref{fig39} and \ref{fig40} show the region of EGO 30 and EGO 32. \citet{kha04} detected eight H$_2$ emission knots in this region, f08-01a-h. Some of the non-detected near-infrared emission knots are located at the position of the stray light marked in Fig.~\ref{fig39} or near the bright star YLW52. The non-detection of these knots may be due to these effects. EGO 30, corresponding to [KGS2004] f08-01a, is a bright arc-like structure. EGO 32a-b are two faint knots that coincide to [KGS2004] f08-01g and h. \citet{gro01} proposed that EGO 30 arises from the interaction of two different jets emanating from two Class I protostars, YLW52 (see Fig.~\ref{fig39}) and YLW15, which is about 10\arcmin \ to the southwest of EGO 30.

\subsection{Discussion and Conclusions}
\begin{figure}
\plotone{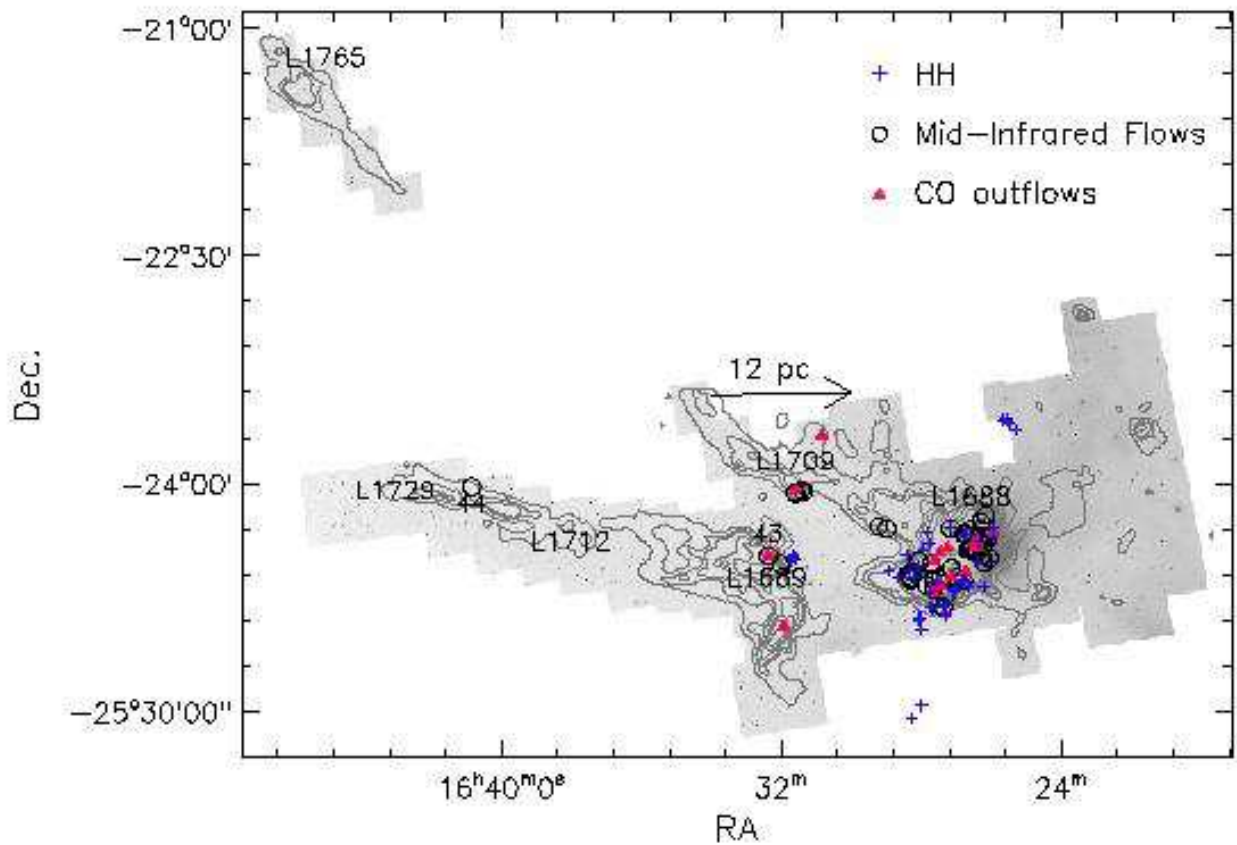}
\caption{The distribution of IRAC mid-infrared outflows, HH objects, and CO outflows in the $\rho$ Ophiuchi molecular cloud. The grey scale image is the IRAC 4.5 \micron \ image. The contours represent antenna temperatures $T^*_{A}$ of 2,4,5,6,7,8,10,12,14,18 and 20K of the $^{13}$CO emission. CO outflows are from \citet{wu04} and HH objects are from {\it Simbad}. The arrow marks the direction of the Upper Scorpius OB association with respect to the Ophiuchus cloud. The distance shown above the arrow is from the center of current figure to the center of Upper-Scorpius OB association region that was defined by \citet{zee99} .\label{fig41}}
\end{figure}

The distribution of mid-infrared outflows in the $\rho$ Ophiuchi molecular cloud, together with that of HH objects \citep{wu02,gom03,phe04} and molecular CO outflows \citep{wu04}, is shown in Fig.~\ref{fig41}. The grey scale image in Fig.~\ref{fig41} is the IRAC 4.5 \micron \ emission and the contours give $^{13}$CO 1-0 emission in the region. It can be seen that mid-infrared outflows, HH objects, and molecular CO outflows roughly have the same spatial distribution and they are in general concentrated in regions with high gas density, particularly in the region of the L1688 dense core. However, difference in the distribution of mid-infrared outflows, HH objects, and molecular CO outflows are apparent in Fig.~\ref{fig41}. First, a substantial number of HH objects, 17 out of 46 that are detected in the $\rho$ Ophiuchi molecular cloud \citep{gww98,wu02,phe04}, are located in regions of $^{13}$CO 1-0 emission less than 2K, while none of mid-infrared outflows and molecular CO outflows are detected in these regions of relatively low gas density. This result may indicate that the excitation of mid-infrared and CO molecular outflows require more dense surrounding medium than HH objects. Secondly, we can see that in the core of the L1688 cloud mid-infrared outflows are distributed in the northern and southern parts of the core while there are several CO molecular outflows detected in the middle of the core. Using outflows as a tracer of current star formation, we can see from Fig.~\ref{fig41} that the core of the L1688 cloud is much more active in current star formation than the core of the L1689 cloud.

\begin{figure}
\epsscale{1.0}
\plotone{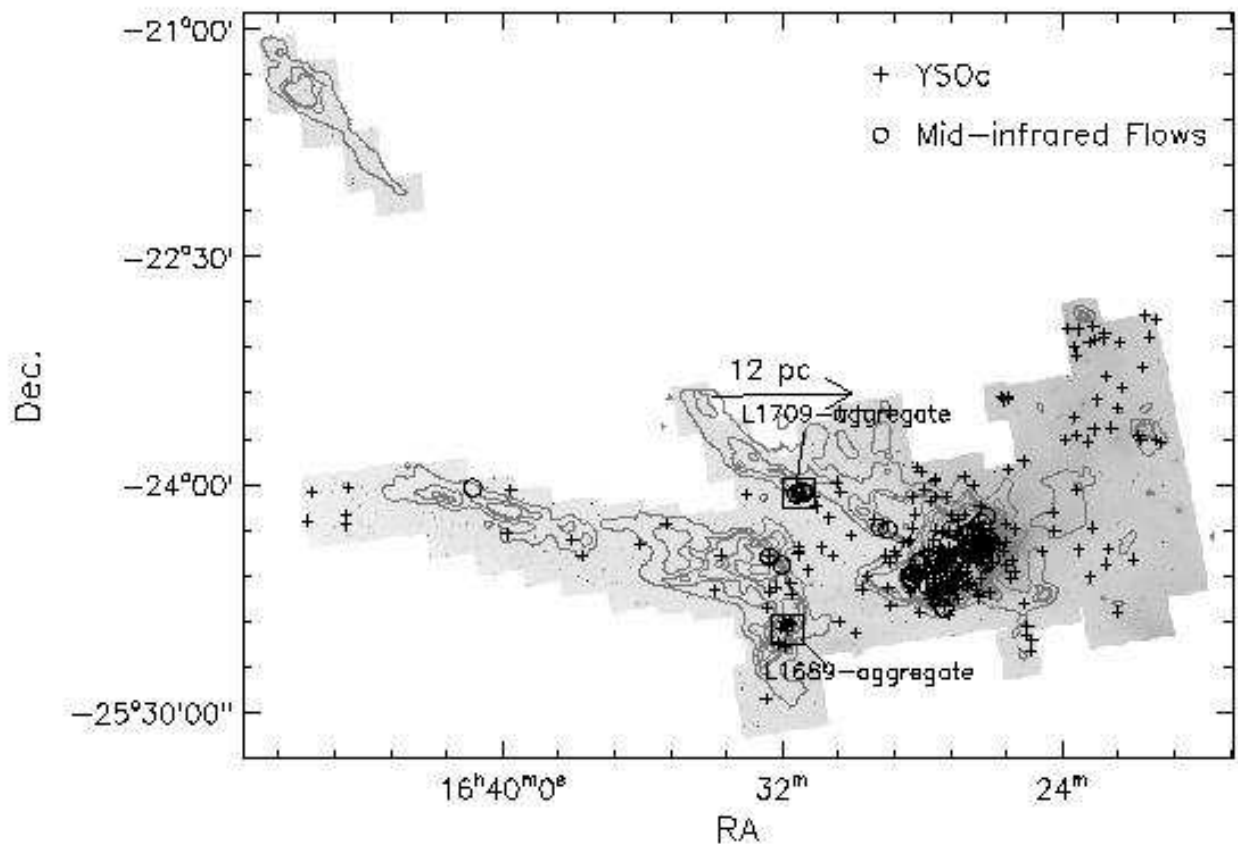}
\caption{The distribution of IRAC mid-infrared outflows compared with that of YSO candidates. Others are the same as in Fig.~\ref{fig41}.\label{fig42}}
\end{figure}

The distribution of YSO candidates in the $\rho$ Ophiuchi molecular cloud is shown in Fig.~\ref{fig42}. YSOs in this region are mainly concentrated in the L1688 cloud core. We note that abundant YSOs are located in the region to the northwest of the L1688 cloud core though there is little dense gas in this region. In fact this part of the $\rho$ Ophiuchi molecular cloud hosts four Class I and two flat-spectrum sources. However, no mid-infrared outflows are detected in this region (see Fig.~\ref{fig42}). The non-detection of mid-infrared outflows in this region may be due to that little dense gas remains in this region.

\begin{figure}
\plotone{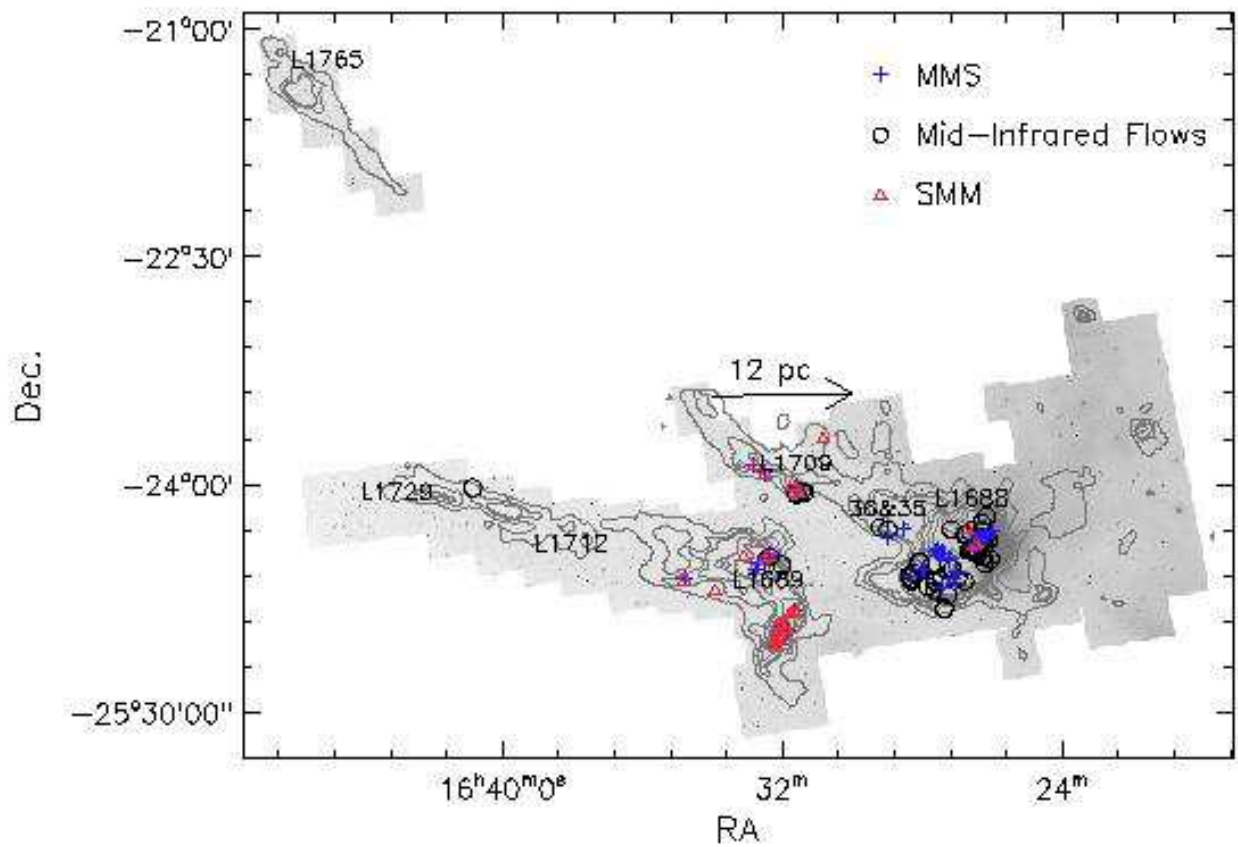}
\caption{The distribution of IRAC mid-infrared outflows compared with that of MMSs and SMMs. MMSs are from \citet{you06} and SMMs are from {\it Simbad}. Others are the same as in Fig.~\ref{fig41}.\label{fig43}}
\end{figure}

Several extensive surveys have been made toward the $\rho$ Ophiuchi molecular cloud for millimeter sources (MMSs) and sub-millimeter sources (SMMs) \citep{wa89,wilson99,vrc02,johnstone04,nwa06,sta06,you06}. The MMSs and SMMs detected in these surveys are shown in Fig.~\ref{fig43}. It can be seen that the detected MMSs and SMMs are exclusively distributed in the L1688 cloud core and the L1689 cloud core. The L1689 cloud core, hosting about 33 MMSs and SMMs, is comparable to the L1688 cloud core, which hosts about 31 MMSs and SMMs. However, we have seen that the L1688 cloud core hosts much more outflows and YSOs than the L1689 cloud core, which may mean that the MMSs and SMMs in the L1689 cloud core are in general in earlier evolutionary stages than those in the L1688 cloud cores and that they are mainly prestellar cores. The distributions of ouflows, YSOs, and MMSs and SMMs  suggest a global star formation history in the $\rho$ Ophiuchi molecular cloud which may be that the star formation in this region took place first in the northwestern part of the $\rho$ Ophiuchi molecular cloud and most dense gas in this region has been dispersed. The L1688 cloud core is now undergoing active star formation while most of the dense cores in the L1689 cloud are still at the prestellar stages. This result is consistent with the suggestion that star formation in Ophiuchus is triggered by ionisation fronts and winds from the Upper Scorpius OB association which is located to the west of the $\rho$ Ophiuchi cloud \citep{lw86,lor89a,nwa06}.

In summary, we detected 13 new mid-infrared outflows in the $\rho$ Ophiuchi molecular cloud that have not been detected previously at other wavelengths. In addition, 31 mid-infrared outflows which correspond to previously detected HH objects or near-infrared emission are detected as well. Seven of these mid-infrared outflow features correspond to previously detected HH objects and 30 to near-infrared emission. Most of the detected mid-infrared outflows are concentrated in the dense core of the L1688 cloud, with only twelve of which locating outside of this core. In combination with the distribution of YSOs and MMSs and SMMs in the $\rho$ Ophiuchi molecular cloud, the distribution of mid-infrared outflows hints a propagation of star formation in the $\rho$ Ophiuchi molecular complex in the direction from northwest to southeast.

\acknowledgements
We wish to thank S. B. Zhang for IDL plotting code which has been uploaded to the web site\footnote{http://code.google.com/p/aicer/}. This work is based on observations made with the {\it Spizer Space Telescope}, which is operated by the Jet Propulsion Laboratory, California Institute of Technology under a contract with NASA. This work was based on data taken by the c2d {\it Spizer Space Telescope} Legacy Programs. This work has made use of the SIMBAD database operated at CDS, Strasbourg, France. We acknowledge the support by NSFC grants 10733030 and 10621303.

\clearpage

\begin{deluxetable}{lccc}
\tabletypesize{\scriptsize}
\tablecolumns{4}
\tablewidth{0pc}
\tablecaption{IRAC Mid-infrared Outflows in Ophiuchus\label{tab1}}
\tablehead{
\colhead{} & \colhead{$\alpha$} & \colhead{$\delta$} & \colhead{Figure}\\
\colhead{Object}    & \colhead{(J2000.0)}  &   \colhead{(J2000.0)} &  \colhead{Reference}}
\startdata
     EGO 04 &   16:26:10.1  &  -24:12:40  & 3,4\\
     EGO 12 &   16:26:18.2  &  -24:15:07  & 3,4\\
     EGO 33 &   16:28:16.2  &  -24:35:03  & 5,6\\
     EGO 35a &   16:28:58.9  &  -24:17:54  & 7,8,9\\
     EGO 35b &   16:28:59.3  &  -24:17:51  & 7,8,9\\
     EGO 35c &   16:28:59.7  &  -24:17:46  & 7,8,9\\
     EGO 35d &   16:28:59.9  &  -24:18:18  & 7,8,9\\
     EGO 35e &   16:29:00.7  &  -24:18:02  & 7,8,9\\
     EGO 35f &   16:29:00.7  &  -24:17:49  & 7,8,9\\
     EGO 35g &   16:29:01.9  &  -24:18:24  & 7,8,9\\
     EGO 35h &   16:29:02.3  &  -24:18:28  & 7,8,9\\
     EGO 35i &   16:29:02.5  &  -24:17:56  & 7,8,9\\
     EGO 35j &   16:29:03.1  &  -24:18:23  & 7,8,9\\
     EGO 36a &   16:29:13.4  &  -24:16:59  & 7,8,9\\
     EGO 36b &   16:29:13.5  &  -24:16:25  & 7,8,9\\
     EGO 36c &   16:29:14.4  &  -24:16:39  & 7,8,9\\
     EGO 36d &   16:29:14.4  &  -24:17:17  & 7,8,9\\
     EGO 36e &   16:29:14.5  &  -24:17:03  & 7,8,9\\
     EGO 36f &   16:29:15.2  &  -24:16:32  & 7,8,9\\
     EGO 36g &   16:29:15.4  &  -24:16:11  & 7,8,9\\
     EGO 36h &   16:29:17.2  &  -24:16:07  & 7,8,9\\
     EGO 37 &   16:31:22.0  &  -24:03:17  & 10,11\\
     EGO 38 &   16:31:25.8  &  -24:02:42  & 10,11\\
     EGO 39 &   16:31:36.0  &  -24:03:51  & 10,11\\
     EGO 40 &   16:31:37.2  &  -24:03:30  & 10,11\\

     EGO 41a &   16:32:01.0  &  -24:31:59  & 12,13\\
     EGO 41b &   16:32:01.2  &  -24:31:54  & 12,13\\
     EGO 41c &   16:32:02.2  &  -24:31:42  & 12,13\\
     EGO 41d &   16:32:03.2  &  -24:32:27  & 12,13\\
     EGO 41e &   16:32:03.2  &  -24:32:19  & 12,13\\
     EGO 41f &   16:32:03.3  &  -24:31:39  & 12,13\\
     EGO 41g &   16:32:04.2  &  -24:31:55  & 12,13\\
     EGO 41h &   16:32:04.2  &  -24:31:35  & 12,13\\
     EGO 41i &   16:32:04.7  &  -24:31:40  & 12,13\\
     EGO 42 &   16:32:11.8  &  -24:30:33  & 12,13\\
     EGO 43a1 &   16:32:19.3  &  -24:28:31  & 13,15\\
     EGO 43a2 &   16:32:19.5  &  -24:28:11  & 13,15\\
     EGO 43a3 &   16:32:21.1  &  -24:28:30  & 13,15\\
     EGO 43a4 &   16:32:21.5  &  -24:28:38  & 13,15\\
     EGO 43a5 &   16:32:22.3  &  -24:28:32  & 13,15\\
     EGO 43a6 &   16:32:23.6  &  -24:28:35  & 13,15\\
     EGO 43b1 &   16:32:27.0  &  -24:28:19  & 13,15\\
     EGO 43b2 &   16:32:27.8  &  -24:28:36  & 13,15\\
     EGO 43b3 &   16:32:28.6  &  -24:28:31  & 13,15\\
     EGO 43b4 &   16:32:29.2  &  -24:28:16  & 13,15\\
     EGO 43b5 &   16:32:30.1  &  -24:28:28  & 13,15\\
     EGO 43b6 &   16:32:33.8  &  -24:29:04  & 13,15\\
     EGO 43b7 &   16:32:34.7  &  -24:28:33  & 13,15\\
     EGO 43b8 &   16:32:35.4  &  -24:29:04  & 13,15\\
     EGO 43b9 &   16:32:36.1  &  -24:28:28  & 13,15\\
     EGO 43b10 &   16:32:36.5  &  -24:28:32  & 13,15\\
     EGO 43b11 &   16:32:36.6  &  -24:28:56  & 13,15\\
     EGO 43c1 &   16:32:44.7  &  -24:29:28  & 13,14\\
     EGO 43c2 &   16:32:45.7  &  -24:29:09  & 13,14\\
     EGO 43c3 &   16:32:47.1  &  -24:29:09  & 13,14\\
     EGO 43c4 &   16:32:47.2  &  -24:28:49  & 13,14\\
     EGO 43c5 &   16:32:48.9  &  -24:29:13  & 13,14\\
     EGO 43d1 &   16:32:55.9  &  -24:28:13  & 13,14\\
     EGO 43d2 &   16:32:57.5  &  -24:28:24  & 13,14\\
     EGO 43d3 &   16:32:58.4  &  -24:28:39  & 13,14\\
     EGO 43d4 &   16:33:01.3  &  -24:28:54  & 13,14\\
     EGO 43d5 &   16:33:02.9  &  -24:28:53  & 13,14\\
     EGO 43d6 &   16:33:04.2  &  -24:28:40  & 13,14\\
     EGO 43d7 &   16:33:07.2  &  -24:28:27  & 13,14\\
     EGO 44 &   16:40:50.6  &  -24:01:21  & 16,17\\

\enddata
\tablecomments{Units of right ascension are hours, minutes, and seconds, and units of declination are degrees, arcminutes, and arcseconds.}
\end{deluxetable}
\clearpage

\begin{deluxetable}{llcr}
\tabletypesize{\scriptsize}
\tablecolumns{4}
\tablewidth{0pc}
\tablecaption{Likely Exciting Sources of the Newly Detected Outflows\label{tab2}}
\tablehead{
\colhead{}    &  \colhead{} &   \colhead{}   &   \colhead{Distance \tablenotemark{c}} \\
\colhead{Object} & \colhead{Exciting Source \tablenotemark{a}} &\colhead{Lada Class \tablenotemark{b}} &\colhead{(arcmin)}}
\startdata
EGO 04 & BKLT J162624-241616 & II & 4.74\\
EGO 12 & BKLT J162624-241616 & II & 1.78\\
EGO 33  & BKLT J162816-243657 & II & 1.88\\
EGO 35  & SR 10, SR 20W, BKLT J162908-241549, SR 20, BKLT J162923-241359 & II, II, II, III, III & 2.84, 16.94\\
EGO 36  & SR 10, SR 20W, BKLT J162908-241549, SR 20, BKLT J162923-241359 & II, II, II, III, III & 1.70, 20.58\\
EGO 37 & GWAYL 4 & Flat & 3.6\\
EGO 38 & GWAYL 4 & Flat & 2.55\\
EGO 39  & L1709 3 & Flat & 0.60\\
EGO 40  & L1709 3 & Flat & 0.70\\
EGO 41  & SSTc2d J163145.81-243909.0, IRAS 16293-2422 & II, 0 & 8.00, 5.50\\
EGO 42  & SSTc2d J163145.81-243909.0, IRAS 16293-2422 & II, 0 & 10.36, 3.00\\
EGO 43 & IRAS 16293-2422 & 0 & 11.35 \tablenotemark{d}\\
EGO 44 & IRAS 16367-2356 & II & 14.90\\
\enddata
\tablenotetext{a}{Source designations used in the table correspond to the following optical or infrared surveys: BKLT is from \citet{bar97}, GWAYL from \citet{gre94}, L1709 from \citet{pad08} and SR from \citet{str49}}.
\tablenotetext{b}{The Lada Class is classified based on the spectral index and bolometric temperature. The criteria are from \citet{gre94} and \citet{chen95}}
\tablenotetext{c}{The distances to the nearest exciting source and to the farthest exciting source if there are more than one possible exciting source.}
\tablenotetext{d}{This is the scale of the outflow.}
\end{deluxetable}
\clearpage

\begin{deluxetable}{lcclr}
\tabletypesize{\scriptsize}
\tablecolumns{5}
\tablewidth{0pc}
\tablecaption{Mid-infrared Counterparts of Known HH Objects and H$_2$ Flows in $\rho$ Ophiuchi\label{tab3}}
\tablehead{
\colhead{}    &  \colhead{$\alpha$} &   \colhead{$\delta$}   &\colhead{Previous} & \colhead{Figure}\\
\colhead{Counterpart} & \colhead{(J2000.0)} & \colhead{(J2000.0)} & \colhead{Identifications\tablenotemark{a}} & \colhead{Reference}}
\startdata
EGO 01a& 16:25:49.8 &-24:22:12  &A\tablenotemark{b}                   &18,19\\
EGO 01b& 16:25:50.4 &-24:22:22  &A\tablenotemark{b}                   &18,19\\
EGO 01c& 16:25:51.3 &-24:22:16  &A\tablenotemark{b}                   &18,19\\
EGO 02& 16:26:03.1 &-24:29:33 &[KGS2004] f10-01b                         &20,21\\
EGO 03& 16:26:06.5 &-24:21:35 &[GSWC2003] 8a                             &22,23\\
EGO 05& 16:26:11.1 &-24:20:33 &[GSWC2003] 8d                             &22,23\\
EGO 06a& 16:26:12.7 &-24:31:13 &[KGS2004] f10-02a                         &20,21\\
EGO 06b& 16:26:13.5 &-24:31:10 &[KGS2004] f10-02b                         &20,21\\
EGO 07a& 16:26:15.9 &-24:22:41 &[GSWC2003] 11a                            &24,25\\
EGO 07b& 16:26:16.4 &-24:22:45 &[GSWC2003] 11b                            &24,25\\
EGO 08& 16:26:16.6 &-24:28:35 &[KGS2004] f10-01e                         &20,21\\
EGO 09& 16:26:16.6 &-24:25:39 &[KGS2004] f10-03                          &24,25\\
EGO 10& 16:26:17.3 &-24:18:05 &[GSWC2003] 9                              &22,23\\
EGO 11& 16:26:17.5 &-24:23:13 &HH 313B\tablenotemark{c}                                   &24,25\\
EGO 13& 16:26:19.0 &-24:23:05 &HH 313A, VLA 1623A, [DMW95] H5                &24,25\\
EGO 14a& 16:26:21.4 &-24:23:31 &[GSWC2003] 14f                            &24,25\\
EGO 14b& 16:26:21.8 &-24:23:37 &[GSWC2003] 14e                            &24,25\\
EGO 14c& 16:26:21.9 &-24:23:26 &[GSWC2003] 14g                            &24,25\\
EGO 15& 16:26:25.3 &-24:27:34 &[KGS2004] f10-01h                         &20,21\\
EGO 16& 16:26:28.4 &-24:27:05 &[KGS2004] f10-01i                         &20,21\\
EGO 17a& 16:26:31.1 &-24:25:14 &[GSWC2003] 20a, [KGS2004] f10-04d, [DMW95] H3 &24,26\\
EGO 17b& 16:26:31.7 &-24:25:17 &[GSWC2003] 20b, [KGS2004] f10-04e           &24,26\\
EGO 17c& 16:26:32.3 &-24:25:19 &[GSWC2003] 20c, [KGS2004] f10-04f           &24,26\\
EGO 17d& 16:26:32.9 &-24:25:22 &[GSWC2003] 20d, [KGS2004] f10-04g, [DMW95] H2 &24,26\\
EGO 17e& 16:26:33.7 &-24:25:29 &[GSWC2003] 20f, [KGS2004] f10-04h, [DMW95] H1 &24,26\\
EGO 18a& 16:26:35.7 &-24:25:27 &[GSWC2003] 21a, [KGS2004] f10-04i           &24,26\\
EGO 18b& 16:26:36.6 &-24:25:39 &[GSWC2003] 21b, [KGS2004] f10-04j           &24,26\\
EGO 18c& 16:26:37.5 &-24:25:40 &[GSWC2003] 21c, [KGS2004] f10-04k           &24,26\\
EGO 18d& 16:26:38.3 &-24:25:46 &[GSWC2003] 21d                            &24,26\\
EGO 18e& 16:26:39.2 &-24:25:54 &[KGS2004] f10-04m                         &24,26\\
EGO 18f& 16:26:40.2 &-24:26:07 &new\tablenotemark{d}                                      &24,26\\
EGO 19& 16:26:38.2 &-24:18:29 &[GSWC2003] 24a                            &27,28\\
EGO 20& 16:26:39.0& -24:40:53 &HH 314                                      &29,30\\
EGO 21a& 16:26:44.4 &-24:20:07 &HH 79b\tablenotemark{c}                                    &27,28\\
EGO 21b1& 16:26:46.4 &-24:20:15 &HH 79a1\tablenotemark{c}                                   &27,28\\
EGO 21b2& 16:26:47.2 &-24:20:12 &HH 79a2\tablenotemark{c}                                   &27,28\\
EGO 22& 16:26:47.5 &-24:38:16 &HH 673, [KGS2004] f04-01a                   &29,30\\
EGO 23& 16:26:59.0 &-24:37:03 &[KGS2004] f04-01b                         &29,30\\
EGO 24a& 16:27:08.1 &-24:33:05 &[KGS2004] f09-01a, [GSWC2003] 3f  &31,32\\
EGO 24b& 16:27:08.4 &-24:33:32 &[KGS2004] f09-01d, [GSWC2003] 3d  &31,32\\
EGO 24c& 16:27:08.6 &-24:33:17 &[KGS2004] f09-01c,               &31,32\\
EGO 24d& 16:27:09.5 &-24:33:41 &[KGS2004] f09-01e, [GSWC2003] 3b  &31,32\\
EGO 25a& 16:27:10.3 &-24:17:45 &[GSWC2003] 25a, HH 711                            &27,28\\
EGO 25b& 16:27:16.4 &-24:43:56 &[GSWC2003] 25b                            &27,28\\
EGO 26a& 16:27:22.3 &-24:49:14 &[KGS2004] f03-02a                         &33,34\\
EGO 26b& 16:27:22.8 &-24:48:53 &[KGS2004] f03-02b, HH 224S                         &33,34\\
EGO 27& 16:27:28.2 &-24:37:56 &[KGS2004] f04-06                          &35,36\\
EGO 28& 16:27:39.6 &-24:41:50 &[KGS2004] f05-01                          &35,36,37,38\\
EGO 29a& 16:27:39.8 &-24:38:09 &[KGS2004] f04-03e, [GSWC2003] 7a &35,36\\
EGO 29b& 16:27:40.2 &-24:37:60 &[KGS2004] f04-03f, [GSWC2003] 7b &35,36\\
EGO 30& 16:27:43.2 &-24:31:51 &[KGS2004] f08-01a               &39,40\\
EGO 31& 16:27:50.4 &-24:40:08 &[KGS2004] f05-03                          &37,38\\
EGO 32a& 16:28:03.9 &-24:30:54 &[KGS2004] f08-01g               &39,40\\
EGO 32b& 16:28:05.2 &-24:30:40 &[KGS2004] f08-01h                         &39,40\\
EGO 34a& 16:28:17.7 &-24:38:09 &[KGS2004] f05-04a                         &5,6\\
EGO 34b& 16:28:19.4 &-24:37:09 &[KGS2004] f05-04b                         &5,6\\
EGO 34c& 16:28:23.1 &-24:36:10 &[KGS2004] f05-04c                         &5,6\\

\enddata
\tablecomments{Units of right ascension are hours, minutes, and seconds, and units of declination are degrees, arcminutes, and arcseconds.}
\tablenotetext{a}{Source designations used in the table correspond to the following optical or infrared surveys: [GSWC2003] is from \citet{gom03}, [KGS2004] from \citet{kha04}, and [DMW95] from \citet{den95}. }
\tablenotetext{b}{This identification is from \citet{luc08} and it is not included in {\it Simbad}}
\tablenotetext{c}{Also detected in the near-infrared and the identifications are designated with HH object numbers.}
\tablenotetext{d}{This source is a new feature which has no counterpart of optical or near-infrared flow.}
\end{deluxetable}
\clearpage

\clearpage

\clearpage
\clearpage

\clearpage

\clearpage
\clearpage

\clearpage
\clearpage

\clearpage

\clearpage

\clearpage
\clearpage

\clearpage

\clearpage

\clearpage

\begin{thebibliography}{}
\bibitem[Allen et al.(2002)]{allen02}Allen, L. E., Myers, P. C., Di Francesco, J., Mathieu, R., Chen, H., \& Young, E. 2002, \apj, 56, 993
\bibitem[Alves de Oliveira \& Casali(2008)]{aoc08} Alves de Oliveira, C., \& Casali, M. 2008, \aap, 485, 155
\bibitem[Andr\'{e} et al.(1990)]{and90} Andr\'{e}, P., Mart\'{i}n-Pintado, J., Despois, D., \& Montmerle, T. 1993, \aap, 236, 180
\bibitem[Andr\'{e} \& Montmerle(1994)]{and94} Andr\'{e}, P., \& Montmerle, T. 1994, \apj, 420, 837
\bibitem[Andr\'{e} et al.(1993)]{and93} Andr\'{e}, P., Ward-Thompson, D., \& Barsony, M. 1993, \apj, 406, 122
\bibitem[Andrews \& Williams(2007)]{andrews07}Andrews, S. M., \& Williams, J. P. 2007, \apj, 671, 1800
\bibitem[Arce et al.(2007)]{arc07}Arce, H. G., et al. 2007, Protostars and Planets V, ed. B. Reipurth, D. Jewitt, \& K. Keil, (Tucson, AZ: Univ. Arizona Press), 245
\bibitem[Bally et al.(2007)]{bal07}Bally, J., Reipurth, B., \& Davis, C. J. 2007, Protostars and Planets V, ed. B. Reipurth, D. Jewitt, \& K. Keil (Tucson, AZ: Univ. Arizona Press), 215
\bibitem[Barsony et al.(1989)]{bar89} Barsony, M., Burton, M. G., Russell, A. P. G., Carlstrom, J. E., \& Garden, R. 1989, \apj, 346, L93
\bibitem[Barsony et al.(1997)]{bar97} Barsony, M., Kenyon, S. J., Lada, E. A., \& Teuben, P. J. 1997, \apjs, 112, 109
\bibitem[Bary et al.(2002)]{bar02} Bary, J. S., Weintraub, D. A., \& Kastner, J. H. 2002, \apj, 576, L73
\bibitem[Bitner et al.(2008)]{bit08} Bitner, M. A., et al. 2008,\apj,688,1326
\bibitem[Bodenheimer(1995)]{bod95}Bodenheime, P. 1995, \araa, 33, 199
\bibitem[Bontemps(1996)]{bontemps96}Bontemps, S. 1996, Ph.D. thesis, Univ. Paris, XI
\bibitem[Bontemps et al.(2001)]{bon01} Bontemps, S., et al. 2001, \aap, 372, 173B
\bibitem[Bouvier \& Appenzeller(1992)]{bou92} Bouvier, J., \& Appenzeller, I. 1992, \aaps, 92, 481
\bibitem[Bussmann et al.(2007)]{bussmann07}Bussmann, R. S., Wong, T. W., Hedden, A. S., Kulesa, C. A., \& Walker, C. K. 2007, \apjl, 657, L33
\bibitem[Caratti o Garatti et al.(2006)]{car06} Caratti o Garatti, A., Giannini, T., Nisini, B., \& Lorenzetti, D. 2006, \aap, 449, 1077
\bibitem[Castets et al.(2001)]{cas01}Castets, A., Ceccarelli, C., Loinard, L., Caux, E., \& Lefloch, B. 2001, \aap, 375, 40
\bibitem[Chen et al.(2007)]{chen07} Chen, X. P., Launhardt, R., \& Henning, Th. 2007, \aap, 475, 277
\bibitem[Chen et al.(1995)]{chen95}Chen, H., Myers, P. C., Ladd, E. F., \& Wood, D. O. S. 1995, \apj, 445, 377
\bibitem[Cyganowski et al.(2008)]{cyg08} Cyganowski, C. J., et al. 2008, \aj, 136, 2391
\bibitem[Davis et al.(2000)]{dav00} Davis, C. J., Dent, W. R. F., Matthews, H. E., Coulson, I. M., \& McCaughrean, M. J. 2000, \mnras, 318, 952
\bibitem[Dent et al.(1995)]{den95} Dent, W. R. F., Matthews, H. E., \& Walther, D. M. 1995, \mnras, 277, 193
\bibitem[Dunham et al.(2008)]{dunham08}Dunham, M. M., Crapsi, A., Evans, N. J., II, Bourke, T. L., Huard, T. L., Myers, P. C., \& Kauffmann, J. 2008, \apjs, 179, 249
\bibitem[Eisl\"{o}ffel et al.(2000)]{eis00}Eisl\"{o}ffel, J., Mundt, R., Ray, T. P., \& Rodr\'{i}guez, L. F. 2000, in Protostars and Planets IV, ed. V. Mannings, A. P. Boss, \& S. S. Russell (Tucson: Univ. Arizona Press), 815
\bibitem[Elias(1978)]{eli78} Elias, J. H. 1978, \apj, 224, 453
\bibitem[Evans et al.(2003)]{eva03} Evans, N. J., II, et al. 2003, \pasp, 115, 965
\bibitem[Evans et al.(2009)]{eva09} Evans, N. J., II, et al. 2009, \apjs, 181, 321
\bibitem[G\'{o}mez et al.(2003)]{gom03} G\'{o}mez, M., Stark, D. P., Whitney, B. A.,\& Churchwell, E. 2003, \aj, 126, 863
\bibitem[G\'{o}mez et al.(1998)]{gww98} G\'{o}mez, M., Whitney, B. A., \& Wood, K. 1998, \aj, 115, 2018
\bibitem[Grasdalen et al.(1973)]{gra73} Grasdalen, G. L., Strom, K. M., \& Strom, S. E. 1973, \apj, 184, L53
\bibitem[Greene et al.(1994)]{gre94} Greene, T. P., Wilking, B. A., Andr\'{e}, P., Young, E. T., \& Lada, C. J. 1994, \apj, 434, 614
\bibitem[Greene \& Young(1992)]{gre92} Greene, T. P., \& Young, E. T. 1992, \apj, 395, 516
\bibitem[Grosso et al.(2001)]{gro01} Grosso, N., Alves, J., Neuh\"{a}user, R., \& Montmerle, T. 2001, \aap, 380, L1
\bibitem[Hirano et al.(2001)]{hir01}Hirano, N., Mikami, H., Umemoto, T., Yamamoto, S., \& Taniguchi, Y. 2001, \apj, 547, 899
\bibitem[Imanishi et al.(2001)]{imanishi01}Imanishi, K., Koyama, K., \& Tsuboi, Y. 2001, \apj, 557, 747
\bibitem[Johnstone et al.(2004)]{johnstone04} Johnstone, D., Francesco, J. D., \& Kirk, H. 2004, \apj, 611, L45
\bibitem[Johnstone et al.(2000)]{johnstone00}Johnstone, D., Wilson, C. D., Moriarty-Schieven, G., Joncas, G., Smith, G., Gregersen, E., Fich, M. 2000, \apj, 545, 327
\bibitem[J\o rgensen et al.(2006)]{jor06} J\o rgensen, J. K., et al. 2006, \apj, 645, 1246
\bibitem[Khanzadyan et al.(2004)]{kha04} Khanzadyan, T., Gredel, R., Smith, M. D., \& Stanke, T. 2004, \aap, 426, 171
\bibitem[Lis et al.(2002)]{lis02}Lis, D. C., Gerin, M., Phillips, T. G., \& Motte, F. 2002, \apj, 569, 322
\bibitem[Loren(1989a)]{lor89a} Loren, R. B. 1989a, \apj, 338, 902
\bibitem[Loren(1989b)]{lor89b} Loren, R. B. 1989b, \apj, 338, 925
\bibitem[Loren \& Wootten(1986)]{lw86} Loren, R. B., \& Wootten, A. 1986, \apj, 306, 142
\bibitem[Lucas et al.(2008)]{luc08} Lucas, P. W., et al. 2008, \mnras, 391, 136
\bibitem[Lynds(1962)]{lyn62} Lynds, B. T. 1962, \apjs, 7, 1L
\bibitem[Mer\'{i}n et al.(2008)]{mer08}Mer\'{i}n, B., et al. 2008, \apjs, 177, 551
\bibitem[Montmerle et al.(1983)]{montmerle83}Montmerle, T., Koch-Miramond, L., Falgarone, E., \& Grindlay, J. E. 1983, \apj, 269, 182
\bibitem[Motte et al.(1998)]{motte98}Motte, F., Andr\'e, P., \& Neri, R. 1998, \aap, 336, 150
\bibitem[Mundy et al.(1992)]{mun92} Mundy, L. G., Wootten, A., Wilking, B. A., Blake, G. A., \& Sargent, A. I. 1992, \apj, 385, 306
\bibitem[Noriega-Crespo et al.(2004)]{nor04} Noriega-Crespo, A., et al. 2004, \apjs, 154, 352
\bibitem[Nutter et al.(2006)]{nwa06} Nutter, D., Ward-Thompson, D., \& Andr\'{e}, P. 2006, \mnras, 368, 1833N
\bibitem[Ozawa et al.(2005)]{ozawa05}Ozawa, H., Grosso, N., \& Montmerle, T. 2005, \aap, 429, 963
\bibitem[Padgett et al.(2008)]{pad08} Padgett, D. L., et al. 2008, \apj, 672,1013
\bibitem[Phelps \& Barsony(2004)]{phe04} Phelps, R. L., \& Barsony, M. 2004, \aj, 127, 420
\bibitem[Preibisch et al.(1998)]{pre98} Preibisch, Th., Guenther, E., Zinnecker, H., Sterzik, M., Frink, S., \& R\"{o}ser, S. 1998, \aap, 333, 619
\bibitem[Reach et al.(2006)]{rea06} Reach, W. T., et al. 2006, \aj, 131, 1479
\bibitem[Reipurth \& Bally(2001)]{rei01} Reipurth, B., \& Bally, J. 2001, \araa, 39, 403
\bibitem[Reipurth et al.(1997)]{rei97} Reipurth, B., Bally, J., \& Devine, D. 1997, \aj, 114, 2708R
\bibitem[Ridge et al.(2006)]{rid06} Ridge, N. A., et al. 2006, \aj, 131, 2921R
\bibitem[Rosenthal et al.(2000)]{ros00} Rosenthal, D., Bertoldi, F., \& Drapatz, S. 2000, \aap, 356, 705
\bibitem[Shang et al.(2007)]{sha07}Shang, H., Li, Z.-Y., \& Hirano, N. 2007, Protostars and Planets V, ed. B. Reipurth, D. Jewitt, \& K. Keil (Tucson, AZ: Univ. Arizona Press), 261
\bibitem[Smith et al.(2005)]{smi05}Smith, M. D., Gredel, R., Khanzadyan, T., \& Stanke, T. 2005, \memsai, 76, 247
\bibitem[Smith et al.(2006)]{smith06}Smith, H. A., Hora, J. L., Marengo, M., \& Pipher, J. L. 2006, \apj, 645, 1264
\bibitem[Smith \& Rosen(2005)]{sr05}Smith, M. D. \& Rosen, A. 2005, \mnras, 357, 1370
\bibitem[Stanke et al.(2006)]{sta06} Stanke, T., Smith, M. D., Gredel, R., \& Khanzadyan, T. 2006, \aap, 447, 609
\bibitem[Stark et al.(2004)]{sta04} Stark, R., et al. 2004, \apj, 608, 341
\bibitem[Struve \& Rudkjobing(1949)]{str49} Struve, O., \& Rudkjobing, M. 1949, \apj, 109, 92
\bibitem[Teixeira et al.(2008)]{tei08}Teixeira, P. S., McCoey, C., Fich, M., \& Lada, C. J. 2008, \mnras, 384, 71
\bibitem[Visser et al.(2002)]{vrc02} Visser, A. E., Richer, J. S., \& Chandler, C. J. 2002,\aj,124,2756
\bibitem[Vrba et al.(1975)]{vrb75} Vrba, F. J., Strom, K. M., Strom, S. E., \& Grasdalen, G. L. 1975, \apj, 197, 77
\bibitem[Wilking et al.(2001)]{wilking01}Wilking, B. A., Bontemps, S., Schuler, R. E., Greene, T. P., \& Andr\'e, P. 2001, \apj, 551, 357
\bibitem[Wilking et al.(2008)]{wil08} Wilking, B. A., Gagn\'{e}, M., \& Allen, L. E. 2008, Astronomical Society of the Pacific, Handbook of Star Forming Regions Vol. II
\bibitem[Wilking et al.(1989)]{wil89} Wilking, B. A., Lada, C. J., \& Young, E. T. 1989, \apj, 340, 823
\bibitem[Wilking et al.(2005)]{wilking05}Wilking, B. A., Meyer, M. R., Robinson, J. G., \& Greene, T. P. 2005, \aj, 130, 1733
\bibitem[Wilking et al.(1987)]{wilking87}Wilking, B. A., Schwartz, R. D., \& Blackwell, J. H. 1987, \aj, 94, 106
\bibitem[Wilking et al.(1997)]{wilking97}Wilking, B. A., Schwartz, R. D., Fanetti, T. M., \& Friel, E. D. 1997, \pasp, 109, 549
\bibitem[Wilson et al.(1999)]{wilson99} Wilson, C. D., et al. 1999, \apj, 513, L139
\bibitem[Wootten(1989)]{wa89}Wootten, A. 1989, \apj, 337, 858
\bibitem[Wootten \& Loren(1987)]{woo87}Wootten, A., \& Loren, R. B. 1987, \apj, 317, 220
\bibitem[Wu et al.(2002)]{wu02} Wu, J., Wang, M., Yang, J., Deng, L., \& Chen, J. 2002, \aj, 123, 1986
\bibitem[Wu et al.(2004)]{wu04} Wu, Y., Wei, Y., Zhao, M., Shi, Y., Yu, W., Qin, S., \& Huang, M. 2004, \aap, 426, 503
\bibitem[Young et al.(1986)]{you86} Young, E. T, Lada, C. J., \& Wilking, B. A. 1986, \apj, 304, L45
\bibitem[Young et al.(2006)]{you06} Young, K. E., et al. 2006, \apj, 644, 326
\bibitem[de Zeeuw et al.(1999)]{zee99} de Zeeuw, P. T., Hoogerwerf, R., \& de Bruijne, J. H. J. 1999, \aj, 117, 354
\end{thebibliography}
\end{document}